\documentclass[12pt]{article}
\usepackage{latexsym}

\let\ssection=\section
\renewcommand{\section}{\setcounter{equation}{0}\ssection}

\newcommand\mathC{\mkern1mu\raise2.2pt\hbox{$\scriptscriptstyle|$}
        {\mkern-7mu\rm C}} 
\newcommand{\mathR}{{\rm I\! R}}         

\newcommand{\scp}[3][]{\ensuremath{\left._{#1}\!\left\langle{#2}\mid{#3}
\right\rangle\!_{#1}\right.}} 

\newcommand\bi{\begin{itemize}}
\newcommand\ei{\end{itemize}}
\newcommand\be{\begin{equation}}
\newcommand\ee{\end{equation}}

\newcommand{\pl}{\ensuremath{{\partial}}}

\begin{document}
\begin{titlepage}

\begin{center}
{\large\bf Reconsidering Relativistic Causality}
\end{center}

\vspace{0.2 truecm}

\begin{center}
        J.~Butterfield \\[10pt] Trinity College \\ Cambridge CB2 1TQ \\ jb56@cam.ac.uk
\end{center}

\begin{center}
       2 August 2007 
\end{center}

	   \begin{center}
	   Forthcoming (after an Appendectomy) in \\
	   {\em International Studies in the Philosophy of Science} volume 21, issue 3, October 
2007
\end{center}

\vspace{0.2 truecm}

\begin{abstract}
I discuss the idea of relativistic causality, i.e. the requirement that causal processes or 
signals can propagate only within the light-cone. After briefly locating this requirement in  
the philosophy of causation, my main aim is to draw philosophers' attention to the fact that 
it is subtle, indeed problematic, in relativistic quantum physics: there are scenarios in 
which it seems to fail. 

I consign to an Appendix two such scenarios, which are familiar to philosophers of physics: 
the pilot-wave approach, and the Newton-Wigner representation. I instead stress two 
unfamiliar scenarios: the Drummond-Hathrell  and  Scharnhorst effects. These effects also 
illustrate a general moral in the philosophy of geometry: that the mathematical  structures, 
especially the metric tensor, that represent geometry get their geometric significance by 
dint of detailed physical arguments.

\end{abstract}

\end{titlepage}

\tableofcontents

\section{Introduction}\label{Intr}
This paper concerns some cases where a relativistic quantum system apparently violates  
relativistic causality, i.e. the requirement that causal processes or signals can travel at 
most as fast as light. This is a large topic, both because there are several apparent cases 
of such spacelike causal processes, and because there are open questions for both physics and 
philosophy.

To save space, I will consign to an Appendix two cases which, though ``heterodox'' traditions 
within physics, are already familiar to philosophers of physics. Namely: (i) the pilot-wave 
approach, in whose relativistic versions the guidance equation and the quantum potential 
yield non-local effects (analogous to their effects in non-relativistic versions); and (ii) 
the Newton-Wigner representation, in which localized states propagate superluminally (and 
which has other non-local aspects, such as the fact that spectral projectors of the position  
operator associated with spacelike related spatial regions do not in general commute).  

I aim instead to draw philosophers' attention to two unfamiliar cases, in which a 
superluminal effect is predicted by an orthodox relativistic quantum theory, viz. quantum 
electrodynamics (QED):\\
\indent (i): the Drummond-Hathrell effect, in which a photon travels at a superluminal speed 
in a curved (general relativistic) spacetime; and \\
\indent (ii): the Scharnhorst effect, in which a photon travels at a superluminal speed 
between two parallel plates in Minkowski spacetime.

Apart from these effects' intrinsic interest, they are worth discussing for three broader 
reasons.\\
\indent (i): Their violation of relativistic causality is very different from that associated 
with the familiar EPR-Bell correlations (since it occurs in a single quantum system, not a 
pair of them). \\
\indent (ii): They illustrate an important moral in the philosophy of geometry: a moral 
which, incidentally, explains away the apparent contradiction in the above claim that a 
photon travels at a superluminal speed. \\
\indent (iii): Although these effects  are uncontroversial in the physics community, our 
present understanding of them is undoubtedly incomplete:  there are open questions, for both 
physics and philosophy,  waiting to be addressed. In particular, there are open 
philosophical, or at least conceptual, questions about: (i) relations between the various 
formulations of relativistic causality, and thereby about which formulations these effects 
violate; and (ii)  how these effects avoid the causal loops, and hence (by the ``bilking 
argument'') the contradictions, that are traditionally meant to follow from superluminal 
processes. So these open questions  are an invitation to the reader for future 
work!\footnote{I confess at the outset that I set aside a recent line of argument  which 
``goes in the opposite direction'' to what follows: i.e. which uses relativistic causality at 
high energies (mostly in the form of an analytic S-matrix) to select low-energy effective 
theories---thereby concluding from effects like those I shall discuss that QED is not 
embeddable in any causal high energy theory. Thanks to Hugh Osborn for this, and for 
referring me to Adams, Arkani-Hamed et al. (2006). That paper also discusses other 
superluminal effects, including one from a brane model---and how it might be detected by tiny 
deviations in the moon's orbit! Shore (2007) includes a response to this line of argument, 
drawing on his work on the Drummond-Hathrell effect: details of which are in Section 
\ref{DHE} below.   So far as I know, this  line of argument awaits philosophers' attention.}

The plan is as follows. First, I place my topic within the general philosophy of causation 
(Section \ref{shad}; this general  discussion is not needed later). In Section 
\ref{troubles}, I discuss different formulations of relativistic causality, emphasising open 
questions about QED in curved spacetime and about avoiding causal loops. Section \ref{moral} 
prepares us for the two effects by discussing a general moral  about the philosophy of 
(chrono)-geometry which they illustrate: that the mathematical  structures, especially the 
metric tensor, that represent geometry get their geometric significance by dint of detailed 
physical arguments. I learnt this moral from Brown (2005), and we need some details of his 
argument for it, in order to understand the two effects. So in Section \ref{moral}, I give 
those details. Once we have them, we will be ready for Section \ref{three}'s punchline: the 
two effects that violate  relativistic causality.\footnote{Though these effects are 
unfamiliar, Brown is not alone in pointing to their philosophical importance. Weinstein 
mentions the Drummond-Hathrell effect as one of several examples, in his recent judicious 
defence of superluminal signalling (2006, pp. 390, 393); and his (1996) urges a moral similar 
to Brown's. What follows is much indebted to them both.} Finally, in the Appendix I discuss 
relativistic causality, or rather is violation, in the pilot-wave approach and the 
Newton-Wigner representation.

\section{In the shadow of causal anti-fundamentalism}\label{shad}
In the history both of physics, and of its relation to philosophy, causation is a perennial 
player, though the role it plays of course changes---dramatically. Similarly in contemporary 
physics, and philosophy of physics: the former seems to be up to its neck in causal talk, and 
accordingly much of the latter focusses on causation---indeed,  an {\em embarras de 
richesse}.

 But I must also admit to embarrassment, in the usual English sense! This arises from my 
being inclined to endorse  {\em causal anti-fundamentalism} in the sense espoused by Norton 
(2003, 2006). Norton denies `that the world has a universal, causal character such as would 
be expressed by a principle of causality that must be implemented in the individual sciences' 
(2006, Section 2). He supports this by surveying  the efforts over the centuries to 
articulate such a  principle, concluding that there is `such a history of  persistent failure 
that only the rashest could possibly expect a viable, factual principle still to emerge' 
(2003, Section 2). Indeed, Norton's summary of his survey provides a helpful thumbnail sketch 
of some roles that causation has played in physics, and these roles' impact on natural 
philosophy, over the last four hundred years. I can best set the scene for my 
discussion---and explain my embarrassment---by an extended quotation:
\begin{quote}
Highlights of this survey include ...  Newton's (1692/93, third letter) insistence that 
unmediated action at a distance is ``…so great an absurdity, that I believe no man, who has 
in philosophical matters a competent faculty of thinking, can ever fall  into it." Yet the 
continued success of Newton's own theory of gravitation, with its lack of any evident 
mediation or transmission time for gravitational action, eventually brought the grudging 
acceptance that this absurdity was not just possible but actual. In the nineteenth century, 
what was required of a process to be causal was stripped of all properties but one, the 
antecedent cause must determine its effect: ``For every event there exists some combination 
of objects or events, some given concurrence of circumstances, positive and negative, the 
occurrence of which is always followed by that phenomenon" (Mill, 1872, Bk. III, Ch.V, §2). 
The advent of quantum mechanics in the early twentieth century established that the world was 
not factually causal in that sense and that, in generic circumstances, the present can at 
best determine probabilities for different futures. So we retracted to a probabilistic notion 
of causation. Yet the principles that we thought governed this probabilistic notion were soon 
proved empirically to be false. For Reichenbach suggested that we could still identify the 
common cause of two events, in this probabilistic setting, by its ability to screen off 
correlations between the events. That too was contradicted by the EPR pairs of quantum 
theory. (Norton, 2006, Section 2)
\end{quote}
Thus Norton rejects in particular the idea that physics provides a reduction of causation, 
for example identifying it with an appropriate transmission of a conserved quantity, 
especially energy or momentum, as in the process theory of causation of Salmon (1984) or Dowe 
(2000). But he does admit (2003, pp. 5-6) that this is currently the most promising candidate 
for a principle of causality.
   
I endorse much of this broad picture; (my disagreements with it will emerge below). But it 
implies that the principal topic in the analysis of causation, on which the evolution of 
physics can shed light, is determinism. And this is ``embarrassing'', for two reasons:\\
\indent (i): In so far as determinism involves a notion of causation, it is a logically weak 
(``thin'') notion. For example: the determining cause is the entire previous state of the 
system, not some logically weaker, because localized, fact or event; (so the determination 
claim is weaker). Also,  determinism does not need, or even support, a necessitarian view of 
laws (not least because determinism  is a feature of a theory, and so only needs a 
theory-relative  notion of law); so that the notion of causation involved in determinism will 
be non-necessitarian. So various philosophers of causation, keen on facts or events as causal 
relata or on causal  necessity,  will find this notion of causation unsatisfactory.  Besides, 
as Norton notes: according to orthodox  quantum theory, even this weak notion fails.  \\
\indent (ii): Furthermore, some proposed weakenings of determinism, such as Reichenbach's 
Principle of the Common Cause, apparently fail in quantum theory.

But fortunately, there remains plenty of useful work to do. In the context of Norton's 
anti-fundamentalism, we can usefully distinguish four kinds of work: this paper will be an 
example of the fourth.

(1): {\em Other sciences}:--- First, even if physics has thus lowered its sights as regards 
causation, the other sciences might nevertheless use a concept of causation that is both 
reasonably strong (``thick'') and in common among them, or many of them. (I think Norton 
could and would agree with this.) Such a situation would again be embarrassing for a 
philosopher of physics, in so far as physics, and its philosophy, apparently has little to 
contribute to pinning down that concept, and fixing its scope and limits. But in any case, 
doing so is of course the aim of many modern philosophers of causation: more power to their 
elbow.\\
\indent But there is also plenty to do {\em within} the philosophy of physics. Here I see 
three kinds of work. 

(2): {\em Causation in classical physics}:--- First of all, one can deny that causation 
within classical physics, or even some small fragment of it such as Newtonian gravitational 
theory, boils down to determinism (as the quotation from Norton suggests). This denial is 
plausible, even if one lacks a fully-fledged theory of causation such as the process theory 
mentioned above. One can simply  consider how Newton's law of universal gravitation, $F = G 
\frac{m_1 m_2}{R^2}$, says that the gravitational force $F$ between two masses $m_1$ and 
$m_2$ depends on the  distance $R$ between them. Though the law is strictly speaking a 
mathematical statement of instantaneous functional dependence, not a causal statement, it has 
usually been taken to state action-at-a-distance---and this has since Newton's time been 
interpreted by many, I daresay most, physicists and philosophers to be a matter of causation.

My own view is that this usual construal is right: but I agree that to defend it would 
ultimately require a theory of causation---a large project which I will duck out of! For this 
paper, I shall simply assume that instantaneous functional dependence can amount to 
instantaneous causation. That is, turning to the relativistic context which is our concern: I 
will allow that functional dependence between values associated with spacelike-related points 
or regions can amount to a violation of relativistic causality---in senses made more precise 
in Section \ref{troubles}.

 I say `can amount to' since we must of course allow that such functional  dependence  in 
some cases merely reflects the joint effects of a common cause, e.g. the values of an 
electromagnetic field at an instant on a sphere around a radiating source. But again, I shall 
duck out of trying to give a general criterion for when spacelike functional dependence 
amounts to spacelike causation, or violation of relativistic causality, rather than merely 
reflecting the joint effects of a common cause. For it will be clear that  my two examples 
involve the former, not merely the latter.\footnote{The remarks in (2) apply equally to the 
Appendix's two examples, the pilot-wave theory and the Newton-Wigner representation. Thus for 
the pilot-wave theory: in the non-relativistic version, the guidance equation is strictly 
speaking a mathematical statement of how the velocity of a particle depends on the position 
of distant particles. But it is usually, and I think rightly, taken to indicate 
action-at-a-distance---and not joint effects of a common cause. A similar remark applies to 
the quantum potential; and there are similar equations, again indicative of 
action-at-a-distance, in relativistic versions. For more discussion, cf. the Appendix.}

(3): {\em Denying the embarrassment}:--- Also, one can reasonably deny Norton's assertions, 
in (i) and (ii) above, that quantum theory gives up determinism and the Principle of the 
Common Cause. As to (i), the pilot-wave approach to quantum theory is deterministic, and 
unrefuted. And as to (ii), there are versions of the Principle of the Common Cause which are 
not only {\em un}refuted by quantum theory's EPR pairs (i.e. violations of the Bell 
inequalities), but provably {\em satisfied} by some rigorous formulations of quantum theory 
(Redei 2002, Redei and Summers 2002, Butterfield 2007). So for both (i) and (ii), there are 
issues that are yet to be settled.

(4): {\em Accepting the embarrassment}:--- Finally, there is plenty to do, even if we accept 
(i) and (ii).  For first,  the topic of determinism is very rich, for philosophical as well 
as technical issues: witness  many of the writings of John Earman (e.g. Earman 1986, 2004, 
2006) . (Also, Section 3 of Norton (2003) describes how determinism can fail even in a simple 
Newtonian example of a ball rolling on a dome; cf. also Norton (2006a).)

 Second, there is plenty to say about {\em relativistic causality}, i.e. the requirement that 
causal processes or signals can propagate only within the light-cone.
Most physicists and philosophers take modern (i.e. post-Einsteinian) physics to endorse this 
requirement wholeheartedly; and I agree that {\em most} modern physics does so. But I submit 
that the requirement is subtler, indeed more problematic, than usually recognized. Not only 
does it have various formulations; also, some examples violate some of the formulations. 

Finally, I should  clarify the relation of these examples to Norton's causal 
anti-fundamentalism. The main point is that my endeavour---scrutinizing relativistic 
causality, and reporting violations of it---is of course compatible with causal 
anti-fundamentalism. Although Norton sees relativistic causality as small beer in comparison 
with the ambitions of causal fundamentalism, he also thereby sees no conflict between modern 
physics' endorsement of relativistic causality and his own causal anti-fundamentalism. 
Indeed, he links the two by claiming that within modern physics, the venerable ``principle of 
causality'' (roughly: that every event has a cause) boils down to exactly the requirement of 
relativistic causality (2006, Section 3). And Norton aside, there is obviously no  conflict 
between  scrutiny of, or even violations of,  relativistic causality, and causal 
anti-fundamentalism.\footnote{Given just his suspicion of causality, Norton could even 
rejoice at my examples violating relativistic causality: he could see them as providing 
another nail in the coffin of (the modern representative of) the principle of causality. But 
as noted in the next paragraph, Norton joins most other philosophers of physics in being 
gung-ho about relativistic causality.}

On the other hand, what follows is opposed to the drift of Norton's discussions, in two ways. 
First:  my endeavour does not positively {\em support} causal anti-fundamentalism. Optimists 
who hope that physics might provide a ``thick'' notion of causation can look to extract such 
a notion from how a precise formulation of relativistic causality treats the notion of 
process or signal. (Below, we shall see some possible places to look.) Second and more 
important: Norton's discussions give the impression that relativistic causality is both 
straightforward to understand, and  endorsed by all post-Einsteinian physics. And of course 
my main message is that  this impression is false.

\section{Formulating relativistic causality}\label{troubles}
There are three general reasons why formulating relativistic causality is difficult. And even 
with precise formulations in hand, it is difficult to relate the theory of Section 
\ref{three}'s effects, viz. QED, to such formulations. I will first  briefly list these 
difficulties (Section \ref{diffies}).  Then I will discuss some precise formulations (Section 
\ref{qnconsensus}), and raise the question of how to avoid causal loops (Section 
\ref{qncontradns}).  But I must admit at the outset that throughout we will see questions 
which I will not address.

\subsection{Difficulties}\label{diffies}
The first reason for difficulty is philosophical. It is not just that there is always a gap, 
and so room for contention, between a formalism and its physical interpretation. There is 
also the more specific trouble  (discussed in Section \ref{shad}, especially (2)) that 
causality, or more commonly in philosophy `causation', is a much-contested concept. In 
particular: when does functional dependence between values associated with spacelike-related 
points or regions  amount to spacelike causation, i.e. a violation of relativistic causality, 
rather than, for example, joint effects of a common cause?

Such questions naturally prompt one to turn to wave physics: to fix on the idea of the 
propagation of a wave (or more generally: a disturbance of a field), and to take relativistic 
causality to prohibit the propagation having a superluminal velocity. But again, there are 
good questions: for example, about the relation to the idea of a signal (e.g. a signal might 
be required to be in some sense controllable, while a disturbance in general need not be). 
Weinstein (2006) is a fine recent review of such questions. But most relevant to us (cf. 
Section \ref{null} et seq.) will be the fact that there are various definitions of the 
velocity of a wave.   

The third reason for difficulty is specific to quantum theory. For it introduces notions and 
considerations independent of classical physics (in particular: independent of the physics of 
waves), in terms of which one can consider formulating relativistic causality. The obvious 
example is the commutativity of spacelike related operators. 

These difficulties no doubt mean that we should expect there to be several different precise 
formulations of relativistic causality; and that we should {\em not} expect their mutual 
logical relations to be clear, or easily ascertained. Indeed, we shall see this in Section 
\ref{qnconsensus}: which will give three formulations drawn from algebraic quantum field 
theory. 

But for us, there is a further problem: even with Section \ref{qnconsensus}'s precise 
formulations in hand, there are difficulties about relating the theory used in Section 
\ref{three}, QED, to them. Indeed, there are two sorts of difficulty here. First, there are 
important open questions about formulating QED on Minkowski spacetime in a rigorous framework 
such as algebraic quantum field theory (`AQFT'): which is what we need for the Scharnhorst 
effect.  Second, there are further difficulties about formulating (i) quantum field theory, 
in particular QED, and (ii) relativistic causality itself, in a curved spacetime: which is 
what we need for the Drummond-Hathrell effect. Here I shall say a bit more about the first 
sort of difficulty, postponing the second to Section \ref{qnconsensus}.

So far as I know, a formulation of QED in the language of AQFT has not yet been 
achieved.\footnote{In this paragraph, I am indebted to Klaas Landsman and Fred Muller for 
correspondence and references.} Of course, there has been impressive work. One major example 
is Steinmann (2000): this book rigorously derives the perturbative formulation of QED, and 
its scattering formalism, from axioms cast in the language of AQFT (2000, Parts II and III).  
But as Steinmann stresses, there are plenty of open questions. And it is not just a matter of 
important topics which are so far unexplored, or threaten to be problematic, within his 
rigorous framework; (among the examples he mentions are gauge invariance and path integrals). 
There is the deeper problem that no one is sure that a rigorous QED exists. Here it is worth 
quoting a passage from his Introduction's summary of the book:
\begin{quote} 
... Next, a description of what we understand under the name of QED is arrived at in a partly 
heuristic way. The resulting definition is not as precise as one might wish. This reflects 
the present state of knowledge. Not only is it not known whether a rigorous theory deserving 
to be called QED exists, it is not even exactly known what ``deserving to be called QED'' 
means. For all we know, there may exist no rigorous QED, or a uniquely defined one, or 
several distinct versions having equal claim to authenticity... Since it is not known how to 
to fashion a coherent, exact, theory ... we set ourselves the humbler task of carrying out 
this program in perturbation  theory, which is the  most intensively studied and best 
understood approximation scheme in QED.  ... (Steinmann, 2000, p. 3)
\end{quote} 
But so much by way of listing difficulties. I turn to

\subsection{Some precise formulations}\label{qnconsensus}
I shall begin with relativistic causality in classical physics, and then consider quantum 
theories in Minkowski spacetime (Section \ref{mink}). Then I consider curved spacetimes 
(Section \ref{cvd}).

\subsubsection{Minkowski spacetime}\label{mink}
 Most presentations of classical relativity theory assert that no ``signal'' or 
``information'' can propagate outside the light cone: there are to be no tachyons. There are 
various more precise formulations of this prohibition, and justifications for it, in the 
physics and philosophy literature; and comparisons of it with other ``locality'' principles: 
for  philosophical  surveys cf. Earman  (1987) and Weinstein (2006, Sections 1-4). But for 
our purposes, we need only recall one way of making it precise: viz. by invoking the 
hyperbolic character of the partial differential equations of a classical field theory. The  
pre-eminent example is Maxwell's equations for electromagnetism on Minkowski spacetime.

\indent For such equations, the state on a spacelike patch $\Sigma$ determines the state on 
the {\em future  domain of dependence} $D^+(\Sigma)$ consisting of the spacetime points $p$ 
such that every past-inextendible causal (i.e. timelike or null) curve through $p$ intersects 
$\Sigma$.  This determination indicates that the state of any field at $p \in D^+(\Sigma)$ 
cannot be influenced by events so far away that an influence from them to $p$ would have to 
be superluminal. We do not need a precise statement of this determination claim, let alone 
its proof. Here it suffices to say that:\\
\indent (i) for an introduction to the initial value problem for classical fields obeying 
hyperbolic equations, cf. e.g. Wald (1984, Section 10.1); (Wald also defines `spacelike  
patch': p. 200);\\
\indent (ii) the proof of this determination claim uses the theory of characteristics for the 
equations concerned: which we will touch on at the end of Section \ref{null};\\
\indent (iii) note for future reference that one similarly defines (a)  the {\em past} domain 
of dependence $D^-(\Sigma)$; and (b) the {\em domain of dependence} as union of the two 
$D(\Sigma) := D^+(\Sigma) \cup D^-(\Sigma)$; and (c) the future etc. domains of dependence of 
an open region $O$ of Minkowski spacetime (rather than a spacelike  patch). 

Turning to relativistic {\em quantum} theories, the situation is more complicated. For the 
interpretative problems of {\em any} quantum theory (relativistic or not, curved spacetime or 
not) about non-locality and the measurement problem are widely regarded as severe, and as 
threatening relativistic causality. In short: non-locality looks like ``spooky 
action-at-a-distance''; and if measurement involves a ``collapse of the wave-packet'', 
perhaps the collapse is superluminal. Besides, relativistic quantum theories raise further 
issues, reflecting the embarrassment, well-known in foundations of physics, that we do not 
have a developed theory of measurement for these theories. For example, a recent line of work 
argues that in such a theory only a restricted class of quantities can be ideally measured, 
on pain of superluminal signaling (Sorkin (1993), Beckman et al. (2001); cf. Weinstein (2006, 
Section 6)). But suppose we set aside these interpretative problems, say by endorsing a 
minimal instrumentalist interpretation, to the effect  that a relativistic quantum theory 
prescribes probabilities for appropriate measurement results.

 Given this supposition, most presentations of relativistic quantum theories in Minkowski 
spacetime agree that the theories incorporate relativistic causality. But still, the 
situation is subtle. The idea has several precise formulations; and in various rigorous 
formalisms, e.g. AQFT, these formulations are precise enough that one can prove e.g. that one 
is logically independent of another.

 As I see matters (and anyway: for the purposes of this paper), the three main formulations 
are as follows: (the first is the analogue of the classical formulation above).

\indent \indent (i): {\em Primitive causality; hyperbolicity}: In a heuristic quantum field 
theory, using the Heisenberg picture, operators indexed by spacetime points are subject to 
Heisenberg equations of motion, while the state is fixed once for all. But these equations  
are hyperbolic, on analogy with classical field theories using hyperbolic dynamical 
equations; this means one can show, at least unrigorously, that for any state, all 
expectation values are determined subluminally, in that the state's restriction to the field 
operators in a region $O$ determines all its expectation values for operators in the  future  
domain of dependence $D^+(\Sigma)$.  In AQFT, this idea is made precise as the {\em Diamond 
Axiom}. AQFT associates to each open bounded region $O$ of Minkowski spacetime, an algebra of 
operators ${\cal A}(O)$. We are to think of the Hermitian elements of ${\cal A}(O)$ as the 
observables for that part of the field system lying in $O$, and so as  measurable by a 
procedure within $O$. We take states as linear expectation functionals $\omega$ on the 
algebras: $\omega: A \in {\cal A}(O) \mapsto \omega(A) \in \mathC$. (We can recover Hilbert 
space representations from this abstract setting, primarily by the GNS construction.)  Then 
the Diamond Axiom  says that ${\cal A}(D(O)) = {\cal A}(O)$. So the idea is: {\em if} $O_1 
\subset D^+(O), O_1 \cap O = \emptyset$, i.e. $O_1$ lies in the top half of the ``diamond'' 
$D^+(O)$, and $A \in {\cal A}(O_1)$, so that we could measure $A$ by a procedure confined to 
$O_1$, {\em then} we could also instead measure $A$ by a procedure confined to $O$. For 
thanks to the hyperbolic time-evolution, ``the facts in $O_1$'' are already determined by 
``the facts in $O$''.

\indent \indent (ii): {\em Spacelike commutativity} (also called {\em micro-causality}): 
Operators associated with spacelike-related regions commute. In heuristic quantum field 
theory, treating fermions requires one to also allow anti-commutation; but in AQFT, one 
distinguishes field algebras and observable algebras, and for the latter imposes only 
spacelike commutativity. Thus one requires: if $O_1, O_2$ are spacelike, then for all $ A_1 
\in {\cal A}(O_1), A_2 \in {\cal A}(O_2): [A_1, A_2] = 0$. The physical idea is of course 
that such spacelike observables should be co-measurable, and so should commute. (Think of how 
in elementary quantum theory, one proves the no-signalling theorem, viz. that a non-selective 
L\"{u}ders rule measurement  of $A$ cannot affect the measurement  probabilities of $B$, 
provided $[A,B] = 0$.)

\indent \indent (iii): {\em Spectrum}: The field system's energy-momentum operator has a 
spectrum (roughly: set of eigenvalues) confined to the future light-cone. Of the three 
conditions (i)-(iii), this is perhaps  the most direct expression of the prohibition of 
spacelike processes.

So we have three different senses of relativistic causality. And indeed, in most relativistic 
quantum theories, and models constructed within them, these three senses (and even others) 
hold good. Besides, there are various arguments connecting these sorts of formulation. For 
example, one can argue for (ii), spacelike commutativity, from an analogue of (i) which one 
might call a ``signal principle'', viz. that ``turning on'' any unitary evolution $U$ of the 
field system within region $O_1$ should leave unaffected the measurement probabilities of any 
observable $A$ associated with a spacelike region $O_2$; along the following  lines.  If 
$O_1, O_2$ are spacelike, then this signal principle requires that for all unitary $U \in 
{\cal A}(O_1)$, all $A \in {\cal A}(O_2)$, and all states $\omega$, $\omega(U A U^*) = 
\omega(A)$. So $UAU^* = A$, i.e. $[U,A] = 0$. Since the unitary operators span the algebras 
concerned, this means that the algebras commute, i.e. for all $A_1 \in {\cal A}(O_1),  A_2 
\in {\cal A}(O_2)$, we have $[A_1, A_2] = 0$.\footnote{I learnt this argument, apparently 
common in the folklore of AQFT, from K. Fredenhagen and S. Summers, in conversation: to whom 
my thanks.}

On the other hand, there are no corresponding rigorous implications between a pair drawn from 
(i) to (iii). That is: truly precise statements of (i) to (iii), in the language of AQFT, are 
logically independent of each other. Indeed, they are independent even in the presence of 
AQFT's other axioms. And more is true: Horuzhy (1990, pp. 19-21) reports that all six of his 
basic axioms of AQFT (three of which are his versions of (i)-(iii)) are independent: for each 
subset comprising five axioms,  there is a model satisfying all five---but not the remaining 
sixth axiom. And although some of these models are very unphysical, others are not: again we 
see that relativistic causality---formulate it as you like!---is a subtle matter in 
relativistic quantum theories.

\subsubsection{Curved spacetime}\label{cvd}
I turn to curved spacetimes. The first thing to say is that in the light of our not having a 
rigorous formulation of QED on Minkowski spacetime (cf. the end of Section \ref{diffies}), 
{\em a fortiori} we lack such a formulation on curved spacetimes. But on the other hand, a 
great deal is now understood about formulating heuristic quantum field theory on such 
spacetimes. I shall sketch the general situation, emphasising the case of a non-interacting 
field (nowadays well understood), and then reporting some recent progress on the (much 
harder) interacting case. In both cases, I shall emphasise ideas we need for our main 
question, to which I will turn at the end of the Section: how to adapt our three precise 
conditions, (i) to (iii), to curved spacetimes.

Broadly speaking, by the mid 1990s quantum field theory on curved spacetime could be 
formulated in as satisfactory a manner as heuristic quantum field theory on Minkowski 
spacetime, subject to three conditions. (This summary is based on Wald (1994).) These 
conditions are: \\
\indent (a): the curved spacetime is fixed, i.e. there is no back-reaction of the field on 
the spacetime geometry; (though the curvature can be non-constant);\\
\indent (b): the field is linear (i.e. not self-interacting);\\
\indent (c): the spacetime is such that the corresponding classical field theory has a  
well-posed initial value problem.

For our purposes, condition (c) prompts two comments. First, `well-posed initial value 
problem' refers to the determination of solutions by initial data, as in the classical 
discussion at the start of Section \ref{mink}. And the most common way to satisfy condition 
(c) is to restrict attention to globally hyperbolic spacetimes. These are spacetimes with a 
{\em Cauchy surface}, i.e. a spacelike slice $\Sigma$ whose domain of dependence $D(\Sigma)$ 
is the whole spacetime. In fact, global hyperbolicity is a strong condition of causal ``good 
behaviour''; it implies that spacetime is foliated by Cauchy surfaces, and implies several 
other causality conditions, including stable causality---which we will return to in Section 
\ref{three}. The reason for this restriction is that the main theorems securing a well-posed 
initial value problem for hyperbolic classical equations assume global hyperbolicity (Wald 
1984, Theorems 10.1.2-3, p. 250.)\footnote{For an approach to quantum field theory on 
non-globally hyperbolic spacetimes, cf. Kay (1992, especially  Section 6).}

Second, we need to explain `the corresponding classical theory'. This phrase indicates a 
standard construction of a heuristic quantum field theory (with e.g. a Fock space of states 
built up from a vacuum) from the solution space of a classical theory. (The construction for 
Minkowski spacetime is in Wald (1994, Sections 3.1-3.2); the adaptation to curved spacetimes 
is in his Section 4.2.) When this construction is given for Minkowski spacetime, but in a 
form that is suitable for generalization to curved spacetimes, one sees that:\\
\indent (i): it involves the choice of a Hilbert space (essentially, of a set of complex 
solutions of the classical theory), with different Hilbert spaces giving unitarily 
inequivalent theories; (the Stone-von Neumann theorem asserting unitary equivalence of 
representations of commutation relations depends on the system being finite-dimensional i.e. 
not a field); and\\
\indent (ii): this freedom of choice is characterized by a choice of a bilinear map (inducing 
an inner product, and so a complex structure); but that\\
\indent (iii): the Poincar\'{e} symmetry of Minkowski spacetime gives a preferred bilinear 
map and so Hilbert space: equivalently, a preferred vacuum with the Hilbert space being the 
Fock space built up from that vacuum; (Wald 1994, pp. 27-29, 39-42).

In a curved spacetime, property (iii) fails, and so one must either: (a): seek some other 
criterion for choosing the bilinear map (or at least a set of them corresponding to a unitary 
equivalence class of representations); or  (b): treat the different choices (and so unitarily 
inequivalent representations) on a par.

In some cases, tactic (a) is sensible. For example, for stationary spacetimes, there is a 
natural unique choice of map; and for a spacetime with a compact Cauchy surface, a condition 
of physical reasonableness for the state (the Hadamard condition, discussed below) constrains 
the bilinear maps so as to fix a unitary equivalence class. But in general, there is no such 
choice and one must opt for (b) (Wald 1994, pp. 58-60). 

This suggests that we should adopt the framework of algebraic quantum theory, which (as 
mentioned in (i) of Section \ref{mink}) takes  abstract algebras of observables as the 
primary notion, with states being linear functionals on them. And indeed, adapting the 
standard construction mentioned above to the  algebraic approach, we find that the (Weyl) 
algebra of observables that naturally arises is independent of the choice of bilinear map 
(Wald 1994, p. 74f.). Furthermore, this approach is at first sight very promising for our own 
topic of precise formulations of relativistic causality, since this approach again associates 
abstract algebras of observables  with open bounded regions of any globally hyperbolic 
spacetime (Wald, 1994 p. 84). Thus we naturally hope to carry over directly to such 
spacetimes the three Minkowski formulations of Section \ref{mink}.

Indeed, there is no problem about the first two conditions, primitive causality and spacelike 
commutativity. The global hyperbolicity assumption prevents any ``funny business'' in the 
causal structure, such as closed causal curves, so that these conditions can be carried over 
word for word: `domain of dependence', `spacelike' etc. now just refer to the curved 
spacetime's structure. 

 Besides, though I have so far confined my summary to the easier, and better understood, case 
of non-interacting fields, the same considerations apply to the interacting case. As we shall 
see shortly, recent formulations of interacting quantum field theory on curved spacetimes use 
the algebraic approach, and again there is nothing to prevent carrying over these two 
conditions intact.

But there is a problem about the third condition, the spectrum condition: though it is a 
problem that has recently been largely solved. In effect, the problem was that no one knew 
how to define the spectrum condition's topic, i.e. the energy-momentum operator, in a curved 
spacetime: all one knew was how to define a class of physically reasonable states that gave a 
well-defined expectation value. But in recent years, the problem has been solved by 
exploiting a mathematical theory, microlocal analysis. Though the details are not needed for 
later discussion, I should summarize them since the problem is much more general, and thus 
its solution much more impressive, than my mentioning just one observable can suggest. 
Indeed, as I understand matters, the solution secures a perturbative formulation of 
interacting heuristic quantum field theory on a globally hyperbolic spacetime that is about 
as precise as those we have for Minkowski spacetime. So, even apart from Section 
\ref{three}'s two effects, this achievement is worth reporting.\footnote{I am very grateful 
to Bob Wald for teaching me what follows. One caveat: to date, the framework is secured for 
scalar fields only. Wald tells me he is very sure it can be extended to Dirac fields, and 
pretty sure for QED: a happy prospect, also for this paper's topic, since it is what we need 
for fully understanding Section \ref{three}'s effects.}

The problem of definability arises from the fact that the energy and momentum of the field 
are encoded in the stress-energy tensor $\hat T$, which involves the square of the quantum 
field ${\hat \phi}$. But $\hat{\phi}$ is a distribution, and the product of distributions at 
a single spacetime point is mathematically undefined; so that some prescription is needed in 
order that  $\hat T$ make sense. Until about 2000, it was not known how to do this 
``directly'', i.e. by enlarging the algebra of observables to include some suitably smeared 
version of  $\hat T$; so one aimed only to characterize a class of physically reasonable 
states $\omega$ for which the expectation value $<{\hat T}>_{\omega}$ was well-defined. This 
was work enough since, in particular, the standard prescription for Minkowski spacetime 
(normal ordering, which corresponds to subtracting off the infinite sum of the zero point 
energies of the oscillators comprising the field) depends on a preferred vacuum---which is 
generally unavailable in a globally hyperbolic spacetime. In fact, there is a compelling 
characterization of such states. Since it builds on Hadamard's work on distributional 
solutions to hyperbolic equations, they are called `Hadamard states'. A bit more precisely: 
one requires the short-distance singularity structure of the two-point function  $<{\hat 
\phi}(x){\hat \phi}(x')>$ to be ``as close as possible'' to the corresponding  structure of 
the two-point function of the Minkowski vacuum (Wald 1994, p. 94). This definition turns out 
to be very successful, as shown by both existence and uniqueness results. That is: (i) any 
globally hyperbolic spacetime has a large class of Hadamard states, and (ii) Hadamard states  
satisfy some natural axioms, and do so uniquely up to a local curvature term; (Wald, 1994: 
pp. 89-95 for (ii), and pp. 95-97 for (i)).

But in recent years, various authors have exploited microlocal analysis so as to achieve the 
original goal (``direct'' in the preceding paragraph). Indeed, they have defined, not just 
the energy-momentum, and  stress-energy operators; but also the other products of field 
operators and their derivatives, and polynomials of such products, and time-ordered products, 
that are crucial in order to formulate the perturbation theory of an interacting quantum 
field theory. I will just gesture at what is involved, with an eye on our interest in the 
spectrum condition. For more detail, cf. Hollands and Wald (2001, 2002), who build on 
previous work, especially by Brunetti, Fredenhagen and collaborators.

The fundamental idea is to use microlocal analysis' definition of the set of directions (at 
each point of the spacetime) in which a distribution is singular: it is called the `wave 
front set' of the distribution. Using this and related ideas, one can define the above 
operators  so as to satisfy appropriate properties, in particular locality and covariance. To 
give the flavour, we need only note the following characterization of Hadamard states in any 
globally hyperbolic spacetime (and thereby of the spectrum condition in Minkowski spacetime): 
the two point function of a Hadamard state has a wave front set consisting of pairs 
comprising: (i)  at any given spacetime point $x$, a future-pointing null vector $k$; and 
(ii) at any other point $x'$ on the null geodesic through $x$ generated by 
parallel-transporting $k$, the corresponding tangent vector, i.e. the parallel-transport of 
$k$ from $x$ to $x'$. (This is, {\em modulo} some technicalities, Radzikowski 1996, Theorem 
5.1.) Hollands and Wald build on this sort of characterization so as to require of their 
local and covariant operators, a generalized microlocal spectrum condition, which is a 
microlocal analogue of the translation invariance of Minkowski spacetime (2001, Definitions 
3.3, 4.1). Thus the spectrum condition, (iii) of Section \ref{mink}, is carried over to 
curved spacetimes.

To sum up Section \ref{qnconsensus}:--- In Section \ref{mink}, we saw three precise 
formulations of relativistic causality for quantum theories on Minkowski spacetime, which 
prescind from interpretative controversies about such notions as causation, or signal, or 
quantum measurement, and which are logically independent. In Section \ref{cvd}, these three 
formulations were adapted to globally hyperbolic spacetimes. But for both kinds of spacetime, 
we do not yet have a rigorous formulation of perturbative interacting QED, in which to study 
the fate of these conditions (and so better understand Section \ref{three}'s effects). But 
there are good prospects for soon getting such formulations.

\subsection{Causal loops?}\label{qncontradns}
Given an apparent violation of relativistic causality, there are various questions to ask. 
First: which of the various precise formulations of relativistic causality is false? (Of 
course, if there are logical connections between them, some may fall as a consequence of 
others, by {\em modus tollens}.) Section \ref{qnconsensus} showed that even when one sets 
aside interpretative problems (as in Section \ref{diffies}), this is a hard technical 
question for either of Section \ref{three}'s effects: hardly a question for philosophers, and 
so a question I must leave to the reader! 

But other questions are more philosophical.  An obvious one is: how does the superluminal 
propagation avoid leading to outright contradictions via causal loops? Although Section 
\ref{three} will not pursue this question, the general strategy  which the investigators of 
the effects have used to address it, is noteworthy---and I can already explain it. Thus the 
investigators of each  example are of course aware of the threat that:\\
\indent (i): superluminal propagation would imply that causal processes could go in a loop 
(technically: closed timelike or null curves); and that: \\
\indent (ii): such loops would yield, by a ``bilking argument'', contradictions. \\
Given this awareness, it is no surprise that they argue that the threat can be avoided: if it 
could not be, the example would be hopeless. But their arguments are worth noting. 

\indent For they do {\em not} adopt the standard philosophical tactic for avoiding the 
threat. That tactic addresses only (ii). Namely, by saying that a ``causal loop'' merely 
implies a severe consistency condition on the state of the system at an initial time, namely 
that it must evolve around the loop back to itself.  Thus, taking the standard philosophical 
example: a contradiction seems to threaten if one can travel back in time and  kill one's 
grandparents before one's parents were conceived. The standard philosophical reply is that 
indeed one could not kill them; but that means, not that time-travel is impossible, but that 
it is severely constrained: any initial state must evolve back to itself; (e.g. Putnam 1962, 
pp. 243-247; and Lewis 1976, pp. 75-79: Earman (1995, pp. 170-188), Berkovitz (2002) and 
Arntzenius and Maudlin (2005, Section 8) are sophisticated discussions of this reply).

\indent Instead, the defences of these examples address (i). That is: they argue that the 
superluminal propagation which the example countenances could {\em not} be exploited to make 
a spacelike zig-zag, ``there and back'', into the causal past of an initial event: a 
refreshing change from the philosophical literature's  repeatedly addressing only the second 
aspect, (ii), by invoking the idea of a consistency constraint.

Another obviously philosophical question is: given superluminal propagation, how should we 
think of the light-cone---what {\em is} its significance, now that it is not the locus of 
light rays, or in a quantum context, of photon propagation? This leads to the more general 
question, which I address in the next Section: how do the mathematical representatives of 
metric structure earn their physical significance? 

\section{Why is this tensor ``read'' by rods and clocks? Brown's moral}\label{moral}
This general question deserves a Section of its own, for two reasons.  First, what I take to 
be the right answer---which I learnt from Brown (2005)---is controversial. So I shall spell 
out what I shall call `Brown's moral', where `moral' connotes both its being controversial, 
and my endorsing it. Second, we need some  details of this moral as preparation for Section 
\ref{three}'s effects; indeed, Brown takes his moral to be illustrated by them (2005, Chapter 
9.4.1, 9.5.2).\footnote{Though the words are mine, this Section  is due to Brown. He urges 
this moral, and related ones, in various passages of his (2005); cf. also his associated 
papers, especially Brown and Pooley (2001, 2006). Since drafting this paper, I have read 
Weinstein's brief but rich (1996). This paper: (i) floats a moral like Brown's, for general 
relativity and kindred theories, and lists some adjacent issues; (ii) illustrates the moral 
with several examples of non-minimal coupling, not just two as in our Section \ref{three}; 
and (iii) discusses coupling in terms of action principles. By the way: though the moral is 
controverted by philosophers, my impression   in discussion with physicists is that they 
endorse it---at least the way I say it!}

Brown's moral is a general doctrine in the philosophy of (chrono)-geometry, though he 
develops it mostly for special relativity, and briefly for general relativity. I will first 
state the moral in general terms (Section \ref{Mgenl}), then report his treatment of it in 
special relativity (Section \ref{Mspecrel}), and then turn to the case we need: general 
relativity (Section \ref{Mgenrel}).

\subsection{The moral in general}\label{Mgenl}
 We can think of the moral as having two aspects, ``negative'' and ``positive''. It will be 
clearer to start with the negative aspect, since the positive  aspect explains it. 
Negatively, the rough idea is that we should not simply postulate that a quantity in a 
physical theory has (chrono)-geometric significance. The point here is not just that it would 
be wrong to infer from a quantity's being {\em called} a metric that it mathematically 
represents (what the theory predicts about) the readings of rods, and-or clocks and-or other 
instruments for measuring lengths and time-intervals. That is obvious enough: after all, a 
quantity might be given an undeserved, even tendentious, name. But also: we should not infer 
from the fact that in the theoretical context, the  quantity is mathematically appropriate  
for representing such behaviour, that it does so. For example, on the Gauss-Riemann 
conception of length as given by line-integrals of $ds = \surd(g_{ij}dx^idx^j)$, the 
symmetric tensor $g_{ij}$ is appropriate. More specifically, for relativity's spatiotemporal 
lengths, $g_{ij}$ is to have Lorentzian signature. But such a tensor might well {\em not} 
represent measured lengths or times. After all, a theory might contain two such tensors, just 
one of which represents such matters. (We shall see such an example in Section \ref{bimet}.)

The reason these inferences are invalid is of course that any physical (chrono)-geometry 
should take rods, clocks and other instruments as composite bodies whose behaviour is 
determined by the laws governing matter. In practice, these bodies are usually very complex 
and so the determination of their behaviour by laws governing their micro-constituents will 
be very complex. But this is not to suggest that a term like `metric tensor' cannot be 
justified. We can indeed write down a idealized (``toy'') model of a rod or clock etc. in 
terms of our theory of matter (nowadays a relativistic and quantum theory), and thereby 
deduce that a certain quantity in our theory---in a relativistic theory, a (0,2) symmetric 
tensor $g_{ij}$ with Lorentzian signature---represents their readings; and thereby earns the 
name `metric'.

We need three further general points about this moral.\\
\indent (1): This is not to say that the quantity must itself be derived, perhaps  as an 
effective or phenomenological aspect of a complex instrument. Our theory can postulate the 
quantity ``on the ground floor'' in its model of the instrument; indeed, most theories do so. 
The point is just that the quantity only earns the name `metric' when we ``close the circle'' 
by exhibiting how instruments' readings display it. (This point is developed in Butterfield 
(2001, Section 2.1.2).)

(2): Nor is it to say that {\em only} by representing the readings of measuring instruments 
for lengths and times (such as rods and clocks), can a quantity have chrono-geometric 
significance. For in both Newtonian and relativistic theories, part of the metric tensor's 
significance is that massive test-particles and light-rays travel along appropriate 
geodesics. (Here, I have put the point in relativistic jargon; and `appropriate' covers 
relativity's timelike/null distinction.)

(3): Brown argues that this moral needs to be stressed, because in recent decades the 
philosophical literature about relativity theory has tended to ignore it. He is especially 
concerned with the moral in special relativity: it is a theme of his Chapters 2-8. For the 
sake of completeness, I will now report his discussion. But this report is not needed for the 
rest of this paper's argument: for that, one can proceed directly to the moral in general 
relativity (Section \ref{Mgenrel}).

\subsection{The moral in special relativity}\label{Mspecrel}
In special relativity, the tensor  $g_{ij}$ get its chrono-geometric significance principally 
through the behaviour of rods and clocks, especially the length contraction and time dilation 
effects.\footnote{One might add: and through massive test-particles and light rays following 
$g_{ij}$'s timelike and null  geodesics, respectively. But I shall set aside this aspect, for 
brevity: compare (2) in Section \ref{Mgenl}.} So as evidence for his moral being ignored, 
Brown describes the current tendency to call them `kinematical effects', where `kinematical' 
is taken to connote `prior to dynamics, and so not needing a dynamical explanation'. To put 
the point  in more philosophical terms: many philosophical commentators on special relativity 
apparently conceive the Minkowski metric as encoding a property (or better: structured family 
of properties) of spacetime that  (a) is intrinsic to it, in the sense that spacetime would 
have the property even in the absence of matter, and (b) suffices to explain the effects.

Brown admits that this tendency has roots, both historical and conceptual. Historically, 
Einstein himself called the opening Sections  of his 1905 paper `Kinematical Part', and he 
called special relativity a `principle' theory as  against a `constructive' one, i.e. as not 
concerned with any detailed mechanisms bringing about length contraction and time dilation. 
And conceptually, the account of length contraction and time dilation based on a spacetime 
diagram with hyperbolae of constant Minkowski interval from a given 
point,\footnote{Originally by Minkowski (1908, pp. 77-78, 81-82), and oft repeated since, 
e.g. Born (1962, pp. 247-249) Torretti (1983, p. 97).}  is undeniably striking. Every student 
feels that the diagram greatly clarifies algebraic derivations based on the Lorentz 
transformations; and that it also makes unmysterious the reciprocity of the effects, i.e. the 
fact that {\em each} of two inertial observers judges the other's rods to be contracted and  
their clocks to be slowed. 

But, says Brown, these roots do not justify the tendency. Einstein himself later admitted 
that the kinematical part of the theory did not pre-empt the need for a dynamical explanation 
of the effects, and accordingly downplayed the idea of special relativity as a principle 
theory. And in order for the account based on a spacetime diagram and hyperbolae to explain 
the effects,  one needs to accept (or to have previously explained) that the primed variables 
do indeed represent the readings of the moving rods and clocks: for only via this fact can 
the diagram's hyperbolae be connected to those readings. And, according to Brown's moral, it 
is of course just this fact that needs  a dynamical explanation.

So much by way of summarising Brown's critique of a current philosophical tendency. Some 
highlights of this critique are at: pp. 22-25, 89-92, 99-102, 129-131, 132-139, and 143-148. 
For example: the critique of the Minkowskian diagrammatic `explanation', i.e. Brown's demand 
for a dynamical explanation of the physical interpretation of the primed variables, is at pp. 
129-131. And on pp. 132-139, Brown argues by analogy with the geometric structure attributed 
to other state spaces in physics, viz. curved configuration manifolds in analytical 
mechanics, the curvature of projective Hilbert space in quantum theory, and 
Carath\'{e}odory's postulates for classical  thermodynamical state-space. In these and 
similar cases, we naturally interpret the geometry not as causing or explaining the system's 
behaviour, but as codifying it: so why not also in spacetime theory? 
 
More positively, Brown gives historical and technical details about the dynamical 
explanations of the effects.  He describes how over the decades several authors, including 
Einstein: (i) have seen the need for such explanations; and (ii) have even  spelt out 
accurately just what such an explanation requires---viz. the Lorentz-covariance of the laws 
responsible for the cohesion of matter, laws which after the 1920s were of course recognized 
to be quantum-theoretic. Some highlights of this positive story are: for (i), Einstein (pp. 
113-114) and Pauli (p. 118); and for (ii), Swann (pp. 119-122) and Bell (pp. 124-126). Brown 
describes how both Swann and Bell realize that:\\
\indent (a): The dynamical explanation does not require one to know what the laws are, but 
only that they are Lorentz-covariant. And:\\
\indent (b): The dynamical explanation does not require a transformation to moving 
coordinates. For example, a Lorentz boost is interpreted as active, mapping a given  solution 
describing a ``stationary'' rod in internal equilibrium, to another solution describing a 
longitudinally contracted rod.

\subsection{The moral in general relativity}\label{Mgenrel}

\subsubsection{How the gravitational field gets its metric significance}\label{313AofMgenrel}
Broadly speaking, Brown's discussion of  general relativity (Chapter 9, and pp. 141-143) 
confirms the moral he gathered from special relativity: that the metric tensor $g_{ij}$ gets 
its chrono-geometrical significance, not by {\em fiat}, but by detailed physical arguments. 
For Section \ref{three}, we need the following details.

All agree that in general relativity, the metric tensor $g_{ij}$ is (or better: represents a 
field that is)  dynamical: it acts and is acted on. They also agree that it is a special 
field since it couples to every other one, and also cannot vanish anywhere in spacetime. Many 
authors go on to say that the metric tensor represents geometry, or spacetime structure, so 
that geometry or spacetime structure acts and is acted on. But Brown resists this. He says 
that the metric tensor  represents primarily the gravitational field, `which interacts with 
every other [field] and thus determines the relative motion of the individual components we 
want to use as rod or clock. Because of that, it admits a metrical interpretation'. (This 
quotation, on p. 160, is from Rovelli: who is one of three distinguished interpreters of 
general relativity whom Brown quotes as kindred spirits.)

Brown supports this position by reviewing some of the physics that underpins this metrical 
interpretation: i.e. the physics that explains why $g_{ij}$ is surveyed by rods and clocks, 
and its null and timelike geodesics are the worldlines of light-rays and massive non-rotating 
test-particles respectively. This review brings out that, with the exception of this last 
case---the worldlines of test-particles---the metrical interpretation depends, not only on 
general relativity's field equations, but also on the {\em strong equivalence principle} 
(SEP). 

This dependence is important for this paper. For it is precisely by violating SEP that 
Section \ref{three}'s effects will have light propagate outside the light-cones defined by 
$g_{ij}$. So it will be worth first spelling out SEP, and seeing how the metrical 
interpretation of $g_{ij}$ uses it. The main point will be that SEP is a conjunction of two 
propositions; and though both are  part of ``textbook general relativity'', only one of them 
(which I will call {\em Universality}) is essential to general relativity---and it will be by 
violating the {\em other} proposition (called {\em Minimal Coupling}) that Section 
\ref{three}'s effects violate SEP, and thereby relativistic causality.

\indent Agreed, what is essential to general relativity is partly a matter of interpretative 
judgment, and partly a purely verbal matter. And unfortunately, there is no sharp consensus 
about how to formulate the equivalence principle; nor about how to make the distinction 
between the weak and strong principles. But I am sure that all general relativists:\\
\indent (i): would accept as reasonable the decomposition of SEP into Universality and 
Minimal Coupling, which Brown articulates (pp. 169-172, citing  Ehlers and Anderson); and \\  
\indent (ii): would accept that only Universality is essential to general relativity; after 
all, there are many articles discussing non-minimal coupling in what the article calls 
`general relativity'.

\subsubsection{The equivalence principles, weak and strong}\label{313BofMgenrel}
So let us start with the {\em weak} equivalence principle. Though I will not need to 
formulate it exactly,\footnote{For  Brown's discussion, cf. pp 25-26 and 161-163. Cf. also 
Norton (1985)  and Ghins and Budden (2001).} the basic idea is that local mechanical 
experiments cannot distinguish gravity and inertia. A bit more precisely: they cannot 
distinguish a homogeneous gravitational field from the inertial effects of uniform 
acceleration. Nor can they distinguish free fall, i.e. motion under gravity but under no 
other forces, from motion subject to no force at all. So this is the idea of ``Einstein's 
elevator'': (which Einstein called ``the happiest thought of my life''). This means in 
particular that test-particles of different masses move in the same way under gravity 
alone---i.e. move identically, given identical initial conditions, and if subject to no other 
force. (``Galileo's law'': two different masses dropped simultaneously from the Tower of Pisa 
fall in identical ways.) Hence the idea of treating gravity as spacetime curvature, in the 
sense of taking freely-falling test-particles to travel along geodesics of a curved 
connection.

On the other hand, the strong equivalence principle, SEP---our main concern---is about  how 
the various non-gravitational forces relate to gravity thus treated; and in particular, how 
they relate to the connection induced by the metric tensor. Again, the formulation varies 
from one author to another. But I will simply follow Brown (and so Ehlers and Anderson), with 
slight modifications. The main point will be that SEP is the conjunction of two propositions. 

The first proposition, Universality, is that the physics of each of the non-gravitational 
forces picks out the same affine connection. More precisely, we envisage that the theory of 
any such force adopts the following framework:\\
\indent (i): The theory is generally covariant. This means, roughly speaking, that it is 
presented in coordinate-independent differential equations for appropriate scalars, vectors 
and tensors representing fields.\\
\indent (ii): The theory invokes an affine connection on spacetime so as to have an 
appropriate coordinate-independent notion of differentiation on fields.\\
Given this framework, Universality then says that all the theories of the non-gravitational 
forces are to invoke the same connection, $\nabla$ say.

\indent This assertion is similar to the basic idea above of the weak equivalence principle, 
for the following reason. Suppose that each such theory asserts that a  test-particle that is 
free---i.e. a particle that is ``small'' enough not to affect what is influencing it, and is 
subject to zero force of the kind in question---travels along a geodesic of the common 
connection $\nabla$. Indeed, in the light of the four-dimensional formulation of Newtonian 
mechanics and special relativity, that is a natural assertion.\footnote{In both these 
theories, there is a distinction between timelike and spacelike curves, and so the theory 
asserts the particle to travel along a timelike geodesic. The main difference is that in 
special relativity, the connection is uniquely determined by (the requirement of 
compatibility with) the metric. We need not consider now the issue whether this assertion 
needs a dynamical or `constructive' explanation of the kind Brown favours. But I will discuss 
this issue in Section \ref{timelike}.} Then Universality makes it very natural to assert that 
in a theory of some or all of these non-gravitational forces,  a test-particle subject to 
{\em none} of these forces---a test-particle that falls freely, i.e. subject only to 
gravity---should also travel along a geodesic of the common connection. After all: if the 
particle did not do so, this would mean that the absence of several forces  yielded, in a 
combined theory of the forces, a motion different from the {\em common} prescription of the 
ingredient theories. So ``agreed votes'' from the ingredient theories would ``cancel one 
another out'' within the combined theory: which would be distinctly odd.

The second proposition, Minimal Coupling, is in effect a bold generalization of this last 
assertion, that in a theory of some or all of the non-gravitational forces, a freely-falling 
test-particle travels along a geodesic of the common connection. This generalization occurs 
in three ways:\\
\indent (i): Minimal Coupling concerns all matter and fields, not just test-particles;\\
\indent (ii): For Minimal Coupling, the matter and fields can be interacting, i.e. subject to 
some or all the non-gravitational forces \\
\indent (iii): Minimal Coupling makes a specific prescription about what laws govern the 
matter and fields in the setting of a curved connection representing gravity. Namely: the 
laws of the corresponding {\em special} relativistic theory are to be valid locally.

It is the third point, (iii), that is the heart of Minimal Coupling. To state it more 
precisely: differential geometry teaches us that the partial derivatives in the differential 
equations of a special relativistic theory implicitly represent the standard flat connection 
of $\mathR^4$; and Minimal Coupling says that the general relativistic laws are given simply 
by replacing these partial derivatives by the curved connection's covariant 
derivatives.\footnote{Partial derivatives are usually represented by a comma, and covariant 
derivatives by a semi-colon. So the semi-colon abbreviates the correction terms, and Minimal 
Coupling is sometimes called the `comma-to-semi-colon' rule.}

\indent Broadly speaking, this prescription means that the transition to the general 
relativistic laws is as simple as it could be, while reducing to the special relativistic 
laws in the case of a flat connection. For a curved connection  means that covariant 
derivatives add ``correction terms'' to  special relativity's familiar partial derivatives. 
But Minimal Coupling says that the general relativistic equations do not add {\em anything 
else}. In particular, they do not include terms proportional to any kind of curvature 
(whether the scalar curvature, or one of the various curvature tensors): which {\em ceteris 
paribus} they might well do, since any such terms would be zero in the setting of special 
relativity's flat connection, and so would not be refuted by the empirical success of the 
special  relativistic theory.

 So Minimal Coupling represents a proposal for simplicity. And  evidently, it is fallible: 
nothing in the framework of general relativity forbids a matter field from coupling to 
spacetime curvature, and so requiring a curvature term in the differential equations that 
govern it. And as announced, we will see examples that violate it, in Section 
\ref{three}.\footnote{{\em Aficionados} know several such: thanks to Steve Adler for 
mentioning the conformal massless Klein-Gordon field.}

\subsubsection{Motion along geodesics}\label{313CofMgenrel}
So much by way of clarifying that the violation of SEP will involve violating Minimal 
Coupling. Returning to Brown's overall moral, we need to extract just two points from his  
review of the physics that underpins this metrical significance of the tensor  $g_{ij}$. 
These concern: (a) motion along timelike geodesics  and (b) light propagation along null 
surfaces. 

\paragraph{Timelike geodesics}\label{timelike} There is one aspect of the metrical 
significance of $g_{ij}$ that is well known to be independent of SEP: viz. the motion of 
massive test-particles. (But again, this aspect will illustrate  Brown's moral.) Thus 
Einstein and other general relativists initially took it as a postulate that freely-falling 
massive non-rotating test-particles followed timelike geodesics of (the curved connection 
determined by) the metric tensor. Brown remarks that this is the analogue for a theory with 
``geometrized gravity'', of the interpretation of Newton's first Law, for Newtonian mechanics 
and special relativity, that he rejects: viz. the interpretation that these theories' 
timelike geodesics `form ruts or grooves in spacetime which somehow guide the free particles 
along their way'; (2005, p. 24; here, `free' of course also excludes gravity; cf. also pp. 
139-143 for references to advocates of this interpretation). But from about 1918 onwards, a 
succession of theorems by Eddington, Einstein and others made it clear that this postulate is 
unnecessary: a massive test-particle must follow such a geodesic, because of the conservation 
of energy-momentum (more precisely: the vanishing of the covariant divergence of the body's 
stress-energy tensor).\footnote{For details, cf. e.g. Misner Thorne and Wheeler (1973, pp. 
471-480), Wald (1984, p. 73), and Geroch and Jang (1975). Brown notes various subtleties 
about these theorems. In particular, although the form of the field equations determines 
$g_{ij}$ to be a (0,2) symmetric tensor, nothing in the equations dictates that $g_{ij}$ 
should have Lorentzian signature.} 

So for Brown, these theorems are like the dynamical explanations of length contraction and 
time dilation in special relativity that he favours. All are genuine physical 
``constructive'' explanations of the chrono-geometric significance of the metric tensor---as 
against the pseudo-explanations that just make a postulate, for example that timelike 
geodesics ``form ruts'' for test-particles. Brown also points out that these theorems are 
limited, but in a way that supports his interpretation. Namely, extended free-falling bodies 
will in general experience tidal gravitational forces, and so will {\em not} follow 
geodesics: underlining the point that it is not ``in the nature'' of freely-falling bodies to 
follow the alleged ruts.

\paragraph{Null surfaces and geodesics}\label{null} 
The propagation of light, or more generally electromagnetic radiation, along null surfaces 
provides an interesting comparison with massive test-particles, Section \ref{timelike} above. 
The situation is similar in that it again illustrates Brown's moral. Thus one can prove from 
Maxwell's equations for electromagnetism on a general relativistic spacetime that 
electromagnetic radiation will propagate along null surfaces. So one does not need to 
postulate that these surfaces ``form ruts'' for light to follow: there are theorems that it 
must do so. But the situation is also dissimilar from Section \ref{timelike}, in that the 
theorems invoke SEP: for it dictates the form that Maxwell's equations take in general 
relativity.

We can see both the similarity and the dissimilarity in the simplest of this kind of theorem: 
viz. where we simplify the description of the light wave, by taking the limit of short 
wavelengths. In this limit, light is described as consisting of rays, with each ray being 
characterized by a curve in spacetime; (and at each point along the curve, an intensity of 
the light, corresponding to the amplitude of the wave---but we need not consider 
intensities). This is called the `geometric optics limit', since geometric optics (as vs. 
wave optics) describes light as composed of such rays.\\
\indent When we take this limit, the direction of the ray is given by the wave-vector (i.e. 
covector, 1-form) $k$ which is the gradient of the phase: the tangent vector to the ray is 
the corresponding contravariant vector $k^i \equiv g^{ij}k_j$. It is straghtforward to show 
from Maxwell's equations, as dictated by the SEP, that:\\
\indent (i): $k$ is a null vector, i.e. $k^2 = 0$; and \\
\indent (ii): the ray is a geodesic, i.e. the ray parallel-transports its own tangent vector: 
$k^i \nabla_i k^j = 0$; (Misner, Thorne and Wheeler 1973, pp. 568-583).

For the purposes of Section \ref{three}, I need to take note of how these theorems can also 
be generalized; i.e. they hold good away from the geometric optics limit. The first point is 
that the propagation of waves is a complex subject, and textbooks of optics, or more 
generally wave physics, sport {\em several} inequivalent notions of the velocity of a wave. 
For example, one (to which we will return) is the {\em phase velocity}: writing the 
wave-vector $ k = (\omega, {\bf k})$ as usual, $v_{\rm{ph}} = \frac{\omega}{\mid {\bf{k}} 
\mid}$.  But for our concerns with causality, it seems to be agreed that the relevant notion 
is {\em wavefront} velocity (also known as: `signal velocity'); which is essentially the 
velocity of the boundary between the regions of zero and non-zero excitation of the field 
concerned. Mathematically, the wavefront is given by the characteristics of the wave equation 
describing the field. So the gist of the general theorems is that the characteristics of 
Maxwell's equations on a general relativistic spacetime are the null surfaces defined by the 
tensor $g_{ij}$; (cf. Friedlander 1975: Theorem 3.2.1).

\indent So to sum up: the moral is as before. It is by a theorem, not by an interpretive 
postulate, that $g_{ij}$  earns the name of `metric'. In particular, SEP implies that 
Maxwell's equations take a form  that makes the ``physical light-cones'' that are defined by 
the (wavefront velocity of/characteristics for) the propagation of light 
coincide exactly with the ``geometric cones'' defined by $g_{ij}(X^i,X^j) = 0$.

\section{Two effects}\label{three}
\subsection{The overall shape of the examples}\label{shape}
The overall shape of the promised examples violating relativistic causality is now clear. At 
the end of Section \ref{moral}, we have seen the conceptual distinction between:\\
\indent (i): the {\em geometric light-cones} defined by the tensor  $g_{ij}$, and \\
\indent (ii): the {\em physical light-cones} that are traversed by electromagnetic radiation; 
(or mathematically, and generalizing from electromagnetism: the characteristics for the wave 
equation governing the field concerned).\\
And we have seen how for electromagnetism in classical  general relativity, SEP makes (i) and 
(ii) coincide.

\indent So we naturally look for a theory that has a regime in which SEP fails, in such a way 
that the physical light-cones turn out to be wider, rather than narrower, than the geometric 
light-cones. That is: the  vectors tangent to the physical  light-cones are to be spacelike, 
rather than timelike---understanding `spacelike' and `timelike' with respect to $g_{ij}$. 
(Here `regime' is physicists' jargon for a certain set of ranges of values of a theory's  
parameters. For example, in fluid mechanics one could  speak of the regime of high density 
and low viscosity. And sometimes, the regime is specified, wholly or in part, by specifying a 
state of the system concerned.)

 More specifically, in terms of the decomposition of SEP (in Section \ref{313BofMgenrel}): we 
look for regimes where Minimal Coupling fails, and  curvature-dependent terms enter the 
action and equations of motion for the electromagnetic field in such a way that 
electromagnetic propagation is described in terms of an {\em effective} metric whose 
light-cones are wider than those of $g_{ij}$. Hence, one talks of `superluminal light': which 
is not a contradiction in terms, since `superluminal' is to be understood as greater than the 
speed $c$ defined by $g_{ij}$.

We consider two such regimes, both arising in quantum electrodynamics (QED): so they concern 
superluminal photons. In fact, they are but two of a family of effects in which vacuum 
polarization affects the propagation of photons, owing to the vacuum being modified by one or 
another external environment.\footnote{A detailed study of photon dispersion and 
birefringence (polarization-dependent
phenomena) is Adler (1971). And in recent years, a framework for unifying these results has 
begun to emerge: e.g. Dittrich and Gies (1998).} But I shall confine myself to these two. The 
first is the Drummond-Hathrell effect, which concerns QED in a general relativistic 
spacetime, i.e. photon propagation in an external gravitational field (Section \ref{DHE}). 
The second is the Scharnhorst effect (Section \ref{SE}), which concerns photon propagation in 
the flat spacetime between two perfectly conducting planar plates.\footnote{Brown's own  
discussion is on p. 165-172. Among other references, he cites Shore (2003) and Liberati et al 
(2002). These and Shore (2003a, 2007), and some of their references, have been my  sources 
for what follows.  I again stress, as in Sections \ref{Intr} and \ref{qncontradns}, that our 
present understanding of these effects, is undoubtedly incomplete---there are plenty of open 
questions hereabouts, for both physicists and philosophers. A vivid illustration of this is 
the recent results about the Drummond-Hathrell effect (Hollowood and Shore 2007, 2007a), 
mentioned in Section \ref{goodbad}.}

\subsubsection{Limitations and alternatives}\label{limaltern}
Before going into details, I should make three other general points about the overall shape 
of these examples. The first two are, I admit, limitations of the examples: about 
observability, and approximations; subsequent Sections will give more details. The third is a 
pointer towards other similar examples, which I will not discuss further.

\paragraph{Observability?}\label{obsy} The word `effect' carries the connotation that one 
could observe it. No such luck, I'm afraid! Both effects are so tiny as to be well beyond 
present observation---and perhaps all future observation. Incidentally: here, `tiny' does not 
mean that all photons travel at a speed greater than $c$ by a tiny amount. Rather (as one 
would expect for a quantum  theory), it means that:\\
\indent (i): there is a  probability distribution for finding the photon to have travelled at 
various speeds; and \\
\indent (ii):  the probability for a photon to be found to have travelled at a speed that is 
greater than $c$, by a large enough margin to be observationally distinguished from $c$, is 
tiny. (In other words: the probability for a photon to be found a measurably large distance 
outside the geometric light-cone is tiny.) 

\paragraph{An artefact of approximations?}\label{artefact} QED is an extraordinarily accurate 
theory, and a long-established one. But it is very complicated, so that calculations within 
it often have to adopt various approximation schemes: and these effects are no exception. We 
shall see that the currently feasible calculations involve various approximations; of which 
one is perhaps especially dubious, as regards the prediction of superluminal photons. Of 
course, our authors stress this; (e.g. Brown 2005, p. 168; Shore 2003, pp. 508, 513, 2003a, 
Sections 3.2, 4.3).
 
\paragraph{The cat out of the bag: bi-metric theories}\label{bimet}  In both these effects, 
the idea of two metrics is ``modest'', in that the second metric is ``merely'' effective. 
That is: it is a structure that helps describe a certain regime of the theory, rather than 
occurring in its fundamental equations describing all regimes. Nor does the structure ``occur 
implicitly'' in the fundamental equations in the sense of  being mathematically determined by 
them, say by being a function of quantities that occur explicitly---and the same function 
regardless of the regime, or a choice of state.\footnote{Incidentally, the emergence in 
relativistic pilot-wave  theories of Lorentz-invariance at the observable quantum level from 
the non-Lorentz-invariant sub-quantum level is an example of such ``implicit occurrence''.}

But once the idea of a theory having two metrics is out of the bag, one naturally speculates! 
Thus a theory might postulate {\em ab initio} two metrics, in the sense of two (0,2) 
symmetric tensors, neither of which mathematically determines the other (so neither is a 
function of the other);  and then the theory might divide between these tensors the various 
roles---or something like the roles---played in general relativity by just the one tensor 
$g_{ij}$.

A second possibility for such a bi-metric theory is a theory with the following three 
features.\\
\indent (1): It postulates a (0,2) symmetric tensor, call it again $g_{ij}$, that is 
fundamental  in that it occurs in the theory's basic equations. \\
\indent (2): But another tensor ${\tilde g}_{ij}$ is a function of the first, ${\tilde 
g}_{ij} = {\tilde g}_{ij}(g_{i'j'})$, where the function concerned is: (i) universal; (ii) 
exact not approximate; (iii) not invertible, or at least not usefully invertible, so that we 
cannot equivalently rewrite the basic equations using ${\tilde g}_{ij}$.\\
\indent (3): And yet our ``critters''---rods, clocks, test-particles and 
light-rays---``display'' ${\tilde g}$ rather than the fundamental tensor $g$.

Indeed, there is a long tradition of such  bi-metric theories. Weinstein (1996) is a fine 
philosopher's introduction to this tradition: launched, like Brown's discussion, from 
consideration of non-minimal coupling (cf. footnote 9). A recent example of the first 
possibility above is Drummond's re-formulation of variable speed of light theories (2001). 
But I shall follow Brown (2005, pp. 172-175) in discussing a recent example of the second 
possibility: the TeVeS theory of Bekenstein. This is a relativistic theory of gravity which 
postulates, in addition to a fundamental (0,2) symmetric tensor $g$, a vector field $U$ 
(which is dynamically constrained to be timelike) and a scalar field $\phi$; and these 
together define a tensor ${\tilde g}$. So `TeVeS' stands for `tensor-vector-scalar'.

The motivation for the theory lies in the tradition of modified Newtonian dynamics (called 
`MOND') begun by Milgrom in the 1980s, to account for the anomalously fast rotation of 
galaxies and clusters without having to invoke dark matter, viz. by making gravity decrease 
more slowly than inverse-square for very large distances. Thus in the TeVeS theory, the 
scalar field $\phi$ makes gravity stronger at large distances; and the vector field $U$ 
enhances gravitational lensing, making the theory empirically adequate to the observations 
that gravitational lensing is stronger than would be expected from the lensing galaxy's 
visible matter: observations which are usually taken to require dark matter.

But I will not need more details of the theory's motivation. What matters for us is that the 
theory defines another tensor ${\tilde g}$ as a function of all three of $g, U$ and $\phi$. 
Roughly speaking, one obtains ${\tilde g}$ by multiplying $g$ in the spacelike directions 
orthogonal to $U$ by a function of $\phi$, and by dividing $g$ parallel to $U$ by the same 
function. From the theory's postulated action, and the resulting equations of motion, one 
shows that the ``critters'' listed above, rods etc., survey ${\tilde g}$, not $g$. So again 
we see Brown's moral: that it is by a detailed physical argument that a tensor---here 
${\tilde g}$, not $g$---earns its chrono-geometric significance.

So much by way of discussing a heterodox, though so far unrefuted, relativistic  theory of 
gravity. Now I turn to proposals for superluminal light propagation in QED in an otherwise 
orthodox relativistic setting: general relativistic for the Drummond-Hathrell effect, and 
special relativistic, though with a preferred rest, for the Scharnhorst effect.

\subsection{The Drummond-Hathrell effect}\label{DHE}
Drummond and Hathrell  (1980) studied the effect on light propagation of vacuum polarization 
in a curved spacetime. Vacuum polarization  gives a photon an effective size characterized by 
the electron's Compton wavelength $\lambda_C = \hbar / mc$ (with $m$ the electron mass). This 
suggests that if the photon propagates in an anisotropic spacetime with typical curvature 
length-scale $L \sim \lambda_C$, its motion might be affected. Here we can already see the 
point made in Section \ref{obsy}, that such an effect on the motion would be tiny; since for 
ordinary astrophysical objects---even the event-horizon of a black hole---the gravitational 
field is weak enough that the curvature length-scale $L$ is vastly larger than 
$\lambda_C$.\footnote{Section \ref{spll?}  will give a numerical estimate for a solar mass 
black hole.} 

Drummond and Hathrell's analysis confirms (subject to three main approximations) the 
suggestion that the photon's motion, in particular its speed, is affected by gravity. That 
is: they deduce an effective action, and corresponding (non-linear) generalizations of 
Maxwell equations (i.e. equations of motion for light), which display an interaction between 
quantized electromagnetism and spacetime curvature. So both the action and the equations 
contain curvature-dependent terms, and violate SEP. Besides, the equations of motion 
imply---again, subject to approximations---that in some scenarios (in particular the 
Friedmann-Robertson-Walker and Schwarzschild spacetimes) light propagates superluminally. I 
will first summarize these results of Drummond and Hathrell, and mention some work by later 
authors (Section \ref{spll?}). Then I will discuss how despite this superluminal propagation, 
causal contradictions can apparently be avoided (Section \ref{avoidcon}).\footnote{My main 
sources are Shore (2003, 2003a); cf. footnote 16.}

\subsubsection{Faster than light?}\label{spll?}
\paragraph{Three approximations}\label{threeapp}
Drummond and Hathrell showed that vacuum polarization implies an effective action for the 
electromagnetic field in a curved spacetime that contains curvature terms and so violates 
SEP. But their derivation is subject to three approximations. They are:\\ 
\indent (i): Only one-loop Feynman diagrams are considered.\\
\indent (ii): Gravity is assumed to be weak, in the sense that the effective action keeps 
only terms of first order in the curvature tensors $R$ (scalar), $R_{ij}$ (Ricci) and 
$R_{ijkl}$ (Riemann). In a bit more detail: this restriction implies that results are valid 
only to the lowest order in the parameter $\lambda^2_C / L^2$, where $L$ is the typical 
curvature length scale; so the results are more accurate, the larger $L$ i.e. the weaker is 
the gravitational field. \\
\indent (iii): The photons are assumed to be low frequency, in the sense that the effective 
action neglects terms involving higher orders in derivatives of the fields.

These approximations are in ascending order of `importance', in the sense of `recalcitrance'! 
That is: \\
\indent As to (i): One-loop  diagrams contribute to the action terms proportional to the fine 
structure constant $\alpha \sim 1/137$, and higher-loop diagrams would contribute terms 
proportional to powers of $\alpha$---which are smaller. So we can expect the one-loop 
approximation to give the dominant effects.  

\indent As to (ii): There is a trade-off here. As mentioned, the effect depends on $L$ being 
comparable with $\lambda_C$; but our results are more accurate the larger $L$ is. 

\indent As to (iii): There are two  reasons why we need to consider high-frequency photons; 
one theoretical and one experimental. The theoretical reason develops the remarks at the end 
of Section \ref{null}. There I reported that for questions about causality, the relevant 
notion of the velocity of a wave (of the {\em several} available!) is the {\em wavefront 
velocity} $v_{\rm{wf}}$: it represents the velocity of the boundary of the region of 
excitation of the field, and  is given mathematically by the characteristics of the field's 
wave equation. I also reported that with SEP,  $v_{\rm{wf}}$ is $c$: i.e. the characteristics 
for the orthodox Maxwell's equations in general relativity are the null hypersurfaces defined 
by $g_{ij}$. But now SEP, and these orthodox equations, are gone, and so we have to ask: what 
is $v_{\rm{wf}}$ for Drummond and Hathrell's non-linear generalization of Maxwell's 
equations? Fortunately, a theorem of Leontovich (from 1972) states that for a large class of 
partial differential equations (including Drummond and Hathrell's equations), the wavefront 
velocity is the infinite frequency limit of the phase velocity. That is: 
\be\label{Leont}
v_{\rm{wf}} \; = \; {\rm{lim}}_{\omega \rightarrow \infty} \; v_{\rm{ph}}(\omega) \; = \; 
{\rm{lim}}_{\omega \rightarrow \infty} \; \frac{\omega}{\mid {\bf{k}} \mid} \; .
\ee   
(The proof is sketched in Shore (2003a, Section 3.2) and (2007, p. 10-11).)
So $v_{\rm{wf}}$ is independent of frequency; but we need to know the high-frequency limit of 
$v_{\rm{ph}}(\omega)$.

\indent The experimental reason relates to the fact that on Drummond and Hathrell's analysis, 
the correction to the photon's speed is $O(\alpha \lambda^2_C / L^2)$, with $L$ the typical 
curvature length scale, as before. Experimentally, this is a ratio of a square of a quantum 
scale $\lambda_C$ to an astrophysical scale, and is therefore minuscule. (In a black hole 
example below, it will be $O(10^{-34})$.) In order to assess whether we could observe this 
correction, Drummond and Hathrell suggest (1980, p. 354) that we make two assumptions:\\
\indent (a): The typical time over which propagation can be followed is characterized by $L$, 
so that the length difference to be observed is given by $\alpha \lambda^2_C / L$.\\
\indent (b): Observability requires this to be $O(\lambda)$ where $\lambda$ is the wavelength 
of the light.\\
These assumptions suggest that observability requires that $\alpha \lambda^2_C / \lambda L \; 
> \; 1$, while our approximations (ii) and (iii) above required respectively that $L \gg 
\lambda_C$ and $\lambda \gg \lambda_C$. Agreed, assumption (b) is unduly pessimistic, since 
modern spectroscopy enables it to be weakened by some six, or even eight, orders of 
magnitude. That is: observability might require only that, say, $\alpha \lambda^2_C / \lambda 
L \; > \; 10^{-8}$. Nevertheless, to observe the effect we would obviously like $\lambda$ as 
small as possible.

So much by way of caveats about the approximations. Turning now to reporting the results, 
there is the proverbial ``good news and bad news'': results giving superluminal propagation, 
and results suggesting that it does not occur. Following Shore, I will report these in order. 

\paragraph{Good news and bad}\label{goodbad}
So far as I know, the main way in which superluminal propagation is derived is by applying to 
Drummond and Hathrell's ``Maxwell's equations'' (derived from their effective action) the 
geometric optics (short wavelength) limit; (cf. Section \ref{null}). When we do this, the 
previous result, that $k$ is null, $k^2 = 0$, is replaced by a more complicated equation. We 
will not need it; but for completeness it is
\be\label{eqnull}
k^2 - \frac{2a}{m^2}R_{ij}k^ik^j + \frac{8b}{m^2}R_{ijmn}k^i k^j a^m a^n \; = \; 0 \; ,
\ee 
where $a$ and $b$ are real constants (of magnitude about 1\% and 0.1\% of $\alpha$), $R_{ij}$ 
is the Ricci tensor, $R_{ijmn}$ is the Riemann tensor and $a$ is the polarization vector (a 
spacelike 4-vector normalized to unity). For us the important  point about eq. \ref{eqnull} 
is that it is homogeneous and quadratic in $k$, and so we can write it in the form
\be\label{introG}
{\tilde g}^{ij}k_ik_j = 0 \; ; \; {\rm {with}} \; {\tilde g}^{ij} \equiv {\tilde 
g}^{ij}(R_{klmn}, a^p) \; .
\ee 
This (frequency-independent) function ${\tilde g}$ of curvature and polarization represents 
an effective metric, as follows.  The tangent  vector to a light ray is now given by $p^i := 
{\tilde g}^{ij}k_j$; i.e. light rays are curves $x^i(s)$ with $dx^i /ds = p^i$.  This 
definition of $p^i$ implies
\be\label{introeff}
({\tilde g}^{-1})_{ij}p^ip^j = {\tilde g}^{ij}k_ik_j \; = \; 0 \; .
\ee
So ${\tilde g}^{-1}$ defines an effective metric: (though  we still raise and lower indices 
with $g_{ij}$). We will from now on write $G$ for ${\tilde g}^{-1}$. So our question is 
whether in some solutions, $G$'s light cones---in Section \ref{shape}'s terms: the physical 
light cones---are wider than the geometric light cones defined by $g_{ij}$. In other words: a 
solution exhibits superluminal light if $p$ is spacelike (with respect to $g_{ij}$)---are 
there such solutions? 

Indeed, Drummond and Hathrell (1980, p. 354; and later authors including Shore) show that the 
Friedmann-Robertson-Walker spacetime, and the Schwarzschild spacetime, are such solutions. 
But as we would expect from the discussion above, the effect is numerically tiny. In 
particular, for classical electromagnetism in the Schwarzschild spacetime, there is a 
geometric optics solution (i.e. a solution of $k^2 = 0, k^i \nabla_i k^j = 0$) describing a 
light ray in a circular orbit with radius $r = 3GM$ (Misner, Thorne and Wheeler (1973) pp. 
672-677); and for one solar mass at the singularity, the quantum correction to the classical 
speed $c$ is $O(10^{-34})$. 

There are also results suggestive of superluminal propagation, for generalizations of the 
Drummond and Hathrell action. In particular, Shore has derived an action containing all 
orders in derivatives (cf. approximation (iii) above): though like Drummond and Hathrell, he 
retains only terms of $O(RFF)$ in curvature $R$ and field strength $F$. And he has shown that 
this action, applied to Bondi-Sachs spacetime (describing gravitational radiation far from 
its source), implies that $v_{\rm{wf}} \equiv v_{\rm{ph}}(\infty)$ is superluminal.   

On the other hand, the ``bad news'': there are indications that these results (even Shore's 
about Bondi-Sachs) are artefacts of the approximations used in deriving them. For a complete 
analysis of high frequency propagation requires one to consider higher-order terms in 
curvature and field strength, not just in derivatives. And there is some mathematical 
evidence---just recently, strong evidence---that these terms will, for high frequencies 
$\omega$, drive $v_{\rm{ph}}(\omega)$, and so $v_{\rm{wf}}$, to c. 

I will not need details about this evidence. But for completeness: the evidence concerns the 
correction to the classical light cone condition $k^2 = 0$ being an integral whose integrand 
contains as a factor a phase roughly of the form 
\be
\exp [- i s^2 \Omega^2(R, \omega) P(R, s)] \; ;
\ee
where $s$ is the integration variable, $R$ is a generic curvature component and  $\Omega(R, 
\omega)$ and $P(R, s)$ are functions (not exactly known) of the gravitational field, and also 
of $\omega$ and $s$ respectively; and where $\Omega^2 \sim \frac{R \omega^2}{m^4}$. So it 
seems likely that whatever the behaviour of $P$ and the other factors in the integrand, high 
frequencies $\omega \rightarrow \infty, \Omega \rightarrow \infty$ and therefore rapid 
variation in the exponent, will drive the integral, i.e. the correction to the classical 
condition $k^2 = 0$, to zero. (For an introduction to these details, cf. Shore (2003: pp. 
516, 518; 2003a, Section 4.3; 2007, pp. 28-29).)

Just recently, this evidence has been much strengthened. Hollowood and Shore (2007, 2007a) 
have shown by means of new techniques that $v_{\rm{wf}} \equiv v_{\rm{ph}}(\infty)$ is always 
$c$. But they also show that micro-causality fails, i.e. commutators of fields at spacelike 
separations do not vanish, but fall off exponentially; so that there is violation of 
relativistic causality in sense (ii) of Section \ref{mink}.

So much by way of reviewing the prospects for superluminal propagation in the 
Drummond-Hathrell effect. Of course, whether (and in what sense) it occurs, our main 
philosophical point---viz. Brown's moral that a tensor earns its chrono-geometric 
significance by dint of detailed physical arguments---is vividly illustrated.  For in this 
effect, the light-cones, and more generally geometric structure, defined by $g_{ij}$ are at 
one remove (though a numerically minuscule one!) from the physical behaviour of light: which 
is instead described by the effective metric $G_{ij}$.

\subsubsection{Avoiding contradictions}\label{avoidcon}
Assuming now that there {\em is} superluminal propagation, I turn to how causal 
contradictions can be avoided. For as announced in (B) of Section \ref{qncontradns}, the 
argument against causal contradictions is not the usual philosophical one, that a closed 
``causal loop'' implies a severe consistency condition which one just presumes the solutions 
in question satisfy. The argument is rather that the kind of superluminal propagation 
envisaged could  not be exploited to produce a causal loop, i.e. a zig-zag, there and back, 
into the causal past of an initial event.

More exactly, Shore makes two points. The first is general (and also made by Drummond and 
Hathrell 1980, p. 353); the second is specific. First, he suggests a zig-zag process from an 
event $A$ to a spacelike event $B$, and then from $B$ to an event $C$ that is spacelike to 
$B$, but in the causal past of $A$, may be expected to require `that the laws of physics 
should be identical in the local frames at different points of spacetime, and that they 
should reduce to their special relativistic forms at the origin of each local frame' (2003, 
p. 511; cf. 2003a, Section 2.1).\footnote{I presume his idea is that only with this can we be 
sure that there can be a process from $B$ to $C$ which is like that from $A$ to $B$. But the 
threatened zig-zag might exploit different processes in its two legs. But nevermind: we will 
see, now and in Section \ref{avoidcon2}, stronger reasons to doubt that there can be such 
zig-zags.} But this requirement is just SEP (cf. Section \ref{313BofMgenrel}): which of 
course fails in the framework of the Drummond-Hathrell effect. 

Second, Shore points out that one can investigate the causal structure given by the effective 
metric $G_{ij}$ by using general notions and results that have been developed for classical 
general relativity, i.e. for the causal structure fixed by the usual metric $g_{ij}$. Thus 
there is a well-known spectrum of properties of causal ``good behaviour''---one of which, 
though strong, is especially relevant to assessing the causal structure fixed by $G_{ij}$. 
This is the concept of {\em stable causality}: I shall not give the exact definition of this; 
(cf. e.g. Hawking and Ellis 1973, p. 198; Geroch and Horowitz 1979, p. 241; Wald 1984, p. 
198). We only need the idea: that a spacetime $(M, g_{ij})$ ($M$ the manifold) is stably 
causal if not only does it lack closed timelike curves, but also the spacetime resulting from 
it by a slight opening out of $g_{ij}$'s light-cones at every point does not have any such 
curves. The idea is of course motivated by wanting causal good behaviour to be robust to 
perturbations. Stable causality also follows from global hyperbolicity, which we saw in 
Section \ref{cvd} to be usually assumed in quantum field theory.

Clearly, stable causality can be applied to our effective metric $G_{ij}$ in two ways:\\
\indent (i): If one knows $(M, g_{ij})$ is stably causal, one can expect that $(M, G_{ij})$ 
has no closed timelike curves, since $G_{ij}$ differs from $g_{ij}$ only by terms of 
$O(\alpha)$; \\
\indent (ii): One can  ask, more ambitiously, whether $(M, G_{ij})$ is itself stably causal; 
so that its causal good behaviour is itself robust to perturbations (in particular  to 
overcoming Section \ref{spll?}'s approximations!). In some cases, this question can be 
answered positively by invoking the following characterization of stable causality. As 
usually stated for $(M, g_{ij})$, it is that a spacetime is stably causal iff it has a 
``global time function'', i.e. a smooth function $f: {\cal M} \rightarrow \mathR$ whose 
gradient is everywhere timelike (Hawking and Ellis 1973, Prop. 6.4.9, p. 198; Wald 1984, 
Theorem 8.2.2, p. 198). But this equivalence of course remains valid for $(M, G_{ij})$ with 
{\em its} definition of `timelike'. So one can be assured that $(M, G_{ij})$ is stably causal 
by finding a global time function $f$.  And indeed, Shore remarks (2003, p. 513; 2003a, 
Section 2.4; 2007, pp. 12, 29) that in the Friedmann-Robertson-Walker spacetime discussed in 
Section \ref{spll?}, the usual global time coordinate is such a function, for $G_{ij}$ no 
less than for $g_{ij}$. So here is a clear case where superluminal light, i.e. the physical 
light cones being wider than the geometric ones, harbours no causal anomalies.

\subsection{The Scharnhorst Effect}\label{SE}
I turn to superluminal light propagation in QED in a special relativistic setting: the 
Scharnhorst effect (Scharnhorst 1990; Barton 1990). Again I will first summarize  results 
(Section \ref{spll?2}). Then I will discuss how causal contradictions can apparently be 
avoided (Section \ref{avoidcon2}).\footnote{My main source is Liberati et al (2002); cf. 
footnote 16.}

\subsubsection{Faster than light?}\label{spll?2}
In 1990, Scharnhorst  and Barton showed that for the vacuum between two infinite perfectly 
conducting plates, QED predicted that photons would have a wavefront velocity $v_{\rm{wf}}$  
larger than $c$. More precisely, $v_{\rm{wf}}$ is enhanced in the direction orthogonal to the 
plane of the plates. Though these derivations were based on approximations, in particular on 
assuming low-frequency photons (cf. (iii) in Section \ref{threeapp}), there is evidence 
(Barton and Scharnhorst 1993, pp. 2040-2044; Scharnhorst 1998, pp. 706-707) that the result 
is not an artefact of the approximations, but a genuine effect,  albeit an unobservably small 
one. (Cf. below for the small size).

But unlike Section \ref{threeapp}, I shall not go into details about the approximations. 
After all (as Barton and Scharnhorst stress: 1993, p. 2044), QED in flat spacetime is much 
better understood than QED in curved spacetime, and the system of two plates has been much 
studied for another striking effect, the Casimir effect: according to which there is a force 
between the plates, even when the quantum electromagnetic field between them is in the vacuum 
state.\footnote{For the history and philosophy of this effect, cf.  Rugh, Zinkernagel and Cao 
(1999).} I will only stress (as Barton, Scharnhorst and Liberati et al. do) that the 
superluminal propagation does not violate Lorentz-invariance of the theory (the action and 
equations of motion which imply the propagation), but  reflects only the 
non-Lorentz-invariance of the vacuum state.

 There are two points here.\\
\indent (i): Suppose a theory obeys a symmetry in the sense that a certain transformation, 
e.g. a spatial rotation or a boost, maps any dynamical solution  to another solution. This by 
no means implies that every solution should be invariant, i.e. mapped onto itself, under the 
transformation: after all, not every solution  of Newtonian mechanics is spherically 
symmetric!   \\
\indent (ii): Agreed, the vacuum state for {\em empty} Minkowski spacetime {\em is} required 
to be Lorentz-invariant since it should ``look the same'' in a translated, rotated or boosted 
frame. But the presence of the plates breaks this symmetry, just as a pervasive 
inertially-moving medium would do: licensing a non-Lorentz-invariant vacuum state.\\
\indent A simple Newtonian example illustrates both these points (given by Liberati et al. 
Section 2.2). A  wavefront of sound spreading from a point source in a fluid at rest in a 
Newtonian spacetime is described as spherical in the rest frame of the fluid, but in a 
Galilean-boosted frame it is described as blunted (due to reduced relative speed) in the 
direction of the boost, and as elongated (increased speed) in the opposite direction. 
 
I turn to numerical details of QED's predictions. Writing $g_{ij}$ for the Minkowski  metric 
(with signature (-,+,+,+)), the effective metric between the plates, i.e. defining (to order 
$\alpha^2$) the physical light-cones of photon propagation, is
\be\label{scheff}
G_{ij} = g_{ij} - \frac{\xi}{1 + \xi}n_i n_j \;\; ;
\ee
where $n^i$ is the unit spacelike vector orthogonal to the plates; (and we again raise and 
lower indices with $g_{ij}$). If $a$ is the distance between the plates, we have 
\be\label{sizexi}
\xi \; = \; 4.36 \times 10^{-32} \; \left( \frac{10^{-6} {\rm{m}}}{a} \right)^4 \; .
\ee 
This is minuscule. But since it is positive, the light-cones of $G_{ij}$ are slightly, albeit 
unobservably, wider in the direction orthogonal to the plates than those of $g_{ij}$: 
superluminal propagation.

\subsubsection{Avoiding contradictions}\label{avoidcon2}
There is of course a large literature about superluminal propagation in a special 
relativistic setting, as part of the yet larger literature on the foundations of special 
relativistic kinematics. Fortunately, Liberati et al. (2002) connect the Scharnhorst effect 
with this literature, of which they give a judicious and detailed discussion; (as of course 
does Brown: 2005, Chapters 2-6). So I am happy to do no more than report some of their main 
points: certainly, I could not do better! I shall report four points: two are general and 
correspond roughly to the first point of Shore's in Section \ref{avoidcon}; two are specific 
to the Scharnhorst effect, and correspond to Shore's second point, about stable causality.

First, Liberati et al. emphasise that Lorentz-invariance does not preclude superluminal 
propagation: a speed $c$ can be invariant without being a maximum signal speed. One sees this 
clearly in the style of derivation of an invariant speed, and of  the Lorentz transformation, 
pioneered by von Ignatowsky in 1910, and repeatedly rediscovered and developed since then. 
These derivations neither assume, nor deduce, that a signal cannot travel faster than $c$. 
(Rather, one deduces that there cannot be a reference frame, or a coordinate system, with a 
relative speed greater than $c$; cf. Liberati et al. 2002, Sections 2.1, 2.3.)

Liberati et al. also emphasise that superluminal propagation does not imply that there can be 
a causal zig-zag from a cause $A$ to a spacelike effect $B$, which is itself the cause of a 
second effect $C$ lying in the causal past of $A$. To sustain this implication, one would 
presumably have to require that `in any given reference frame, the only criterion for saying 
that an event is the cause of another one [is] the time ordering in that frame'. Which is 
surely false. Although `there are no precise definitions of [cause and effect] ... the 
criteria used to establish that $e_1$ is a cause of $e_2$ are based on considerations ... 
about the so-called ``arrow of time'' '(2002, Section 3.1).

But whatever the vagaries of the notions of cause and effect, there is obviously no threat of 
a contradiction provided that with respect to one particular reference frame, all 
superluminal propagation is forward in time.\footnote{Drummond and Hathrell also make this 
point (1980, p. 353). One might add: consider such propagation, or even instantaneous 
action-at-a-distance, in a Newtonian spacetime.} And turning now to the Scharnhorst effect, 
it is straightforward to check that this is so, with respect to the rest frame of the plates. 
Indeed, here we connect with the notions of causal good behaviour, in particular stable 
causality, studied in classical general relativity and invoked by Shore, as we saw in Section 
\ref{avoidcon}. Thus we need only consider the spacetime $\mathR^4$ containing the two 
infinite plates, distance $a$ apart, and equipped with the usual Minkowski metric outside the 
plates, and the effective metric eq. \ref{scheff} inside. It is obvious that the time 
coordinate $t$ of the rest frame of the plates has $\nabla t$ everywhere non-zero and 
timelike. So the spacetime is stably causal: not only are there no closed timelike curves; 
but also none could arise by a slight widening of the cones throughout the spacetime (i.e. 
between the plates, a further widening from $G_{ij}$). Obviously, a similar argument will 
secure stable causality for scenarios with more than one pair of plates, at least if they do 
not move relative to one another.

Finally, what about pairs of plates in relative motion? Liberati et al. consider various 
scenarios, arguing that a causal contradiction will not arise. Then they end by suggesting 
that the general threat of contradiction should be analyzed using Hawking's Chronological 
Protection conjecture. The idea of the conjecture is that if a spacetime is causally 
well-behaved ``early on'', it cannot become badly-behaved later. More precisely: a region of 
closed timelike curves which does not extend indefinitely into the past must have a ``first'' 
closed null curve, at which---Hawking argues---uncontrollable singularities will occur, 
implying the breakdown of quantum field theory on curved spacetime, and the need for some 
sort of quantum theory of gravity; (cf. Earman 1995, pp. 188-193).  Applying this idea to a 
spacetime in which early on, several pairs of plates are well separated, and individually, 
stably causal (cf. above), we infer that if some scheme for the plates' later motion seems to 
yield a causal contradiction, then, as Liberati et al. put it: `causal paradoxes are the 
least of your worries since you are automatically driven into a regime where Planck-scale 
quantum gravity holds sway' (2002, Section 3.2.4).

\section{Conclusion}
By way of concluding this paper, let me briefly list three of my main claims:--- \\
\indent (i): In some QED scenarios, relativistic causality is apparently violated.\\
\indent (ii): These scenarios raise open questions, not just about how to define relativistic 
causality, and how to avoid causal contradictions (in more interesting ways than by saying 
`there is a severe consistency constraint'), but about the much wider  question, of the 
future of relativistic quantum physics. \\
\indent (iii): Philosophically, these scenarios illustrate Brown's moral that the 
mathematical representatives of geometry get their geometrical significance by dint of 
detailed physical arguments.\\

{\em Acknowledgements}:-- For comments, conversation and correspondence, I am very grateful 
to: audiences in Dubrovnik, Oxford and Cambridge; two referees, and Bill Demopoulos, Ian 
Drummond, Gordon Fleming, John Norton, Brian Pitts, Graham Shore; and especially to Steve 
Adler, Harvey Brown, Dennis Lehmkuhl, Bob Wald and Steven Weinstein.

\section{Appendix: Two familiar examples}\label{2examples}
This Appendix reports the two examples of quantum-theoretic violations of relativistic 
causality which are most familiar to philosophers of physics: the pilot-wave approach, and 
the Newton-Wigner representation. I shall say much more about the former.

Both examples concern Minkowski spacetime. So one naturally asks which of Section 
\ref{mink}'s precise formulations are violated. But this is a subtle, and even controversial, 
question, since the relations of these examples' formalisms to the local field operators that 
are those formulations' topic, are indirect: and in some respects, unknown or controversial. 
So here again there are questions I cannot pursue: it must suffice that the references below 
are the place to begin  finding the answers.

\subsection{The pilot-wave}\label{p-w}
 There are various pilot-wave approaches to relativistic quantum theory. But I shall sketch 
one well-developed approach which is strongly analogous to the best-known pilot-wave approach 
to non-relativistic quantum theory. As an example of violating relativistic causality, it is 
in a sense only an example ``by courtesy'': for it takes the relativistic light-cone 
structure as a ``merely emergent'' or phenomenological description. For this reason, and also 
because of its being analogous to Newtonian action-at-a-distance, this example provides an 
interesting comparison with the others. I shall:  begin with the pilot-wave approach to 
non-relativistic quantum theory (Section \ref{nrpw}); then discuss action-at-a-distance 
within it, and the Newtonian analogy (Section \ref{aaad}); and finally turn to the 
relativistic case (Section \ref{rpw}).

\subsubsection{The non-relativistic case}\label{nrpw}
Recall that the system comprises both a wave and one or more point-particles. Let us begin 
with the wave and how it evolves over time. The wave is a complex-valued function $\psi$ on 
configuration space ${\cal Q}$ (e.g. ${\cal Q} = \mathR^{3N}$ for $N$ spinless particles in 
euclidean space), which  always evolves by the Schr\"{o}dinger equation. The Schr\"{o}dinger 
equation is local in the mathematical sense: roughly, the evolution depends on $\psi$ and its 
spatial derivatives but not on differences of $\psi$ at different points in $\cal Q$. But it 
is non-local in the physical senses that:\\
\indent (i): it is defined on configuration space, not real space; and \\
\indent (ii): wave-functions of even a single particle propagate instantaneously: if at time 
$t = 0$ a wave-function $\psi(0)$ has compact spatial support (i.e. is non-zero only in a 
compact spatial region), then at all later times $t$, no matter how small, $\psi(t)$ is 
non-zero throughout all space.\\
\indent Despite the continuous deterministic Schr\"{o}dinger evolution, an analysis of 
measurement processes demonstrates an effective collapse of the wave function, nowadays often 
called `decoherence': which explains the instrumental success of orthodox textbooks' 
notorious projection postulate.\footnote{More precisely, it explains it, once allied to the 
pilot-wave theory's invoking particle positions to provide definite events. That decoherence 
alone is not enough to solve the measurement problem is argued by e.g. Bub (1997, pp. 
221-223, 231-232, 236) and Adler (2003).}

\indent Each point-particle has a continuous  trajectory which is determined by the 
wave-function according to the guidance equation. We need to note three features of the 
guidance equation:---\\
\indent (i): {\em Classical analogues}: This equation is natural. Indeed, it follows from the 
orthodox probability current for the Schr\"{o}dinger equation found (for the one-particle 
case) in most textbooks. It is also the obvious wave-mechanical analogue of a central 
equation of classical Hamilton-Jacobi theory, viz. $p = \frac{\pl S}{\pl q}$. More generally, 
much of the pilot-wave approach bears comparison with Hamilton-Jacobi  theory. In particular, 
many quantum effects are due to the presence (in the analogue of the classical 
Hamilton-Jacobi equation) of an extra potential term dependent on the wave-function, viz. the 
quantum potential $U$. In the simplest one-particle case, $U := - 
\frac{\hbar^2}{2m}\frac{\nabla^2 R}{R}$; where  $R$ is given by the polar decomposition of 
the wave-function, i.e. $\psi({\bf q},t) = R({\bf q},t)\exp(iS({\bf q},t)/\hbar)$.  \\
\indent (ii): {\em Probability and equivariance}: The pilot-wave approach recovers the 
orthodox quantum probabilities by averaging over particle position using $\mid \psi \mid^2$ 
as the probability density. This is the Born rule, understood non-instrumentalistically; i.e. 
understood with probabilities like those of classical statistical mechanics. Besides: taken 
together with the Schr\"{o}dinger equation, the guidance equation implies that this 
probabilistic interpretation of the wave-function, viz. that $\mid \psi \mid^2$ is the 
position probability density, is  preserved over time. \\
\indent (iii): {\em Non-locality}:  The guidance equation also implies that in a 
multi-particle system, the motion of each particle is determined in part by the simultaneous 
actual positions of the other particles. For the guidance equation says that the particle's 
momentum is the gradient of the phase $S$ of the wave-function at the  point in configuration 
space corresponding to all these actual positions. To be precise, for the $i$th particle, and 
the $N$ actual positions ${\bf q}_j, j = 1, \dots, N$: ${\bf p}_i \equiv m{\dot {\bf q}_i} = 
\; \nabla_i S \mid_{{\bf q}_1, ... , \; {\bf q}_N}$.\\
\indent \indent Non-locality is also evident in the quantum potential for many particle 
systems. Thus for two spinless particles, so that $\psi({\bf q}_1,{\bf q}_2,t) = R({\bf 
q}_1,{\bf q}_2,t)\exp(iS({\bf q}_1,{\bf q}_2,t)/\hbar)$, the quantum potential is $U({\bf 
q}_1,{\bf q}_2,t) = - \frac{\hbar^2}{2m}\frac{1}{R}(\nabla^2_1 R + \nabla^2_2 R)$; and only 
for the special case of product states $\psi({\bf q}_1,{\bf q}_2,t) = \psi_1({\bf 
q}_1,t)\psi_2({\bf q}_2,t)$, is the quantum potential a sum, $U = U_1({\bf q}_1,t) + U_2({\bf 
q}_2,t)$.\\
\indent \indent On the other hand, this non-locality cannot be exploited to send a signal in 
the sense of affecting the statistics of distant experiments---provided the system is in 
``quantum equilibrium'', i.e. provided that $\mid \psi \mid^2$ is indeed the position 
probability density. That is, the orthodox quantum no-signalling theorem is recovered by 
averaging over particle positions with $\mid \psi \mid^2$.

\subsubsection{Action-at-a-distance?}\label{aaad}
In Section \ref{nrpw}'s summary, two features look like action-at-a-distance: the 
instantaneous spreading of wave-functions, and the non-local guidance equation. (In this 
Subsection, I set aside the quantum potential, discussion of which would be similar to that 
of the guidance equation.)

\indent The first is undeniable, but also unsurprising. For this is a feature of {\em 
orthodox} non-relativistic quantum theory which the pilot-wave  approach simply 
inherits---and which one naturally expects to disappear in a relativistic theory, not least 
because in  a relativistic setting such superluminal propagation seems to threaten causal 
contradictions. But cf. Section \ref{rpw} and Section \ref{NW}.

 But we need to pause here over the second, the guidance equation. I agree that it is natural 
to take it as asserting action-at-a-distance; (though as just noted, this ``action'' in an 
individual process could only be used to send a signal if the system was in quantum 
dis-equilibrium). But I should register that this can be (and has been) resisted---for 
reasons that apply equally to the more familiar case of Newton's law of universal 
gravitation, $F = G \frac{m_1 m_2}{R^2}$, mentioned in (2) of Section \ref{shad}. So I shall 
explain both the natural construal and the reasons for resistance, for both cases.

For both the pilot-wave and Newton's law, the value of a quantity is a function of the 
simultaneous value of another quantity. For gravity, the force is determined by the 
simultaneous value of distance (or position of the other mass); for the pilot-wave, the 
momentum (or velocity) is determined by the simultaneous value of many positions. (Agreed, 
there are other differences which are crucial for physical calculations and explanations: 
above all, that the size of the effects of Newtonian gravity drops off with distance, while 
that of the pilot-wave need not.) In both cases, it is very natural to take the mathematical 
statement of functional dependence as a causal statement, asserting action-at-a-distance. And 
it is natural to support this causal construal by appealing to counterfactuals. Thus it is 
natural to say that  in a Newtonian world, a counterfactual like `If the centre of the Sun 
were now, say, 10,000 km away from where it actually is, the Sun's gravitational pull on the 
Earth would {\em now} be in a slightly different direction' is  true. And the second 
occurrence of `now' (rather than `eight minutes from now') suggests instantaneous causation. 
In particular, within contemporary philosophy of causation: an advocate of the counterfactual 
analysis of causation (Lewis 1973) would take this counterfactual to indicate instantaneous 
causation.

 And similarly for the guidance equation. For example, for the two particles in an EPR-Bell 
experiment: it is natural to say that a counterfactual like `If the L-particle were now in a 
different position, the velocity  of the R-particle would now be different' is true, and 
indicates causation. 

But I admit that one can resist this; in fact Dickson (1998, Section 9.4, pp. 196-208) does 
so. His overall position is reminiscent of Norton's causal anti-fundamentalism (Section 
\ref{shad}). He is cautious about causal judgments, and suspicious of the counterfactual 
analysis of causation. Like Norton, he suggests that under determinism, the most we can 
safely say is that the {\em total} present state is the effect of earlier states: he is wary 
of logically weaker, localized, facts or events as causal relata.\footnote{Beware of a false 
move. We could not really avoid action-at-a-distance, by (i) allowing such localized facts or 
events as causal relata, and then (ii) saying that the various present localized facts or 
events (e.g. in the Newtonian case: $m_1$'s position, and the force on $m_2$) are joint 
effects of a common cause, viz. the total state at any earlier time. For one can consider a 
very recent time, and the state at that time of a very distant region---say, the positions 
and velocities of bodies in that region; and so get ``as near as makes no difference'' to 
instantaneous causation, viz. the causal ``contribution''  from the distant bodies to the 
nearby present fact or event.} 

In more detail: Dickson agrees that in everyday life we tend to interpret counterfactuals as 
non-backtracking, i.e. to take a counterfactual supposition about a state of affairs at time 
$t$ to preserve  most of the past of $t$. And this makes us endorse the counterfactuals 
above, for we think along the following lines: if the Sun were now in a different place, 
nevertheless its past and the Earth's past would be as it actually was until very recently, 
and so the Earth would be very nearly where it actually is, and so would indeed feel the 
Sun's pull in a slightly different direction. But, says Dickson, it is unclear what this 
interpretative tendency---apparently conventional and so alterable---has to do with 
causation. Thus he and the counterfactual analysis agree that if instead we take 
counterfactuals as backtracking, then for a deterministic theory like Newtonian gravity, a 
counterfactual supposition about the  present implies differences from actuality indefinitely 
far into the past: differences which could in general grow as we go further into the past 
(Lewis 1973a: pp. 75). But Dickson sees this as a sign---not of how backtracking 
counterfactuals are irrelevant to causation---but of how poorly we understand causation, in a 
deterministic world no less than (and perhaps even more than!) in a indeterministic one.   

So much by way of:\\
\indent (i) reporting  the traditional, natural construal of the guidance equation (and 
Newton's law of gravitation) as involving instantaneous causation; and \\
\indent(ii) registering that one can nevertheless resist this. \\
My own view is that the traditional construal is right: but as I admitted in (2) of Section 
\ref{shad}, to defend this view would be a large project in the philosophy of causation, 
which I duck out of. (And as I said there, I also do not have a general criterion of when 
spacelike functional dependence amounts to spacelike causation (a violation of relativistic 
causality), rather than merely reflecting the joint effects of a common cause. But I think it 
is clear that all my examples involve the former, not merely the latter.)

So to assess whether the pilot-wave approach violates relativistic causality, I turn to ...

\subsubsection{The relativistic case}\label{rpw}
I turn to the adaptation of the pilot-wave approach in Section \ref{nrpw} to relativity. 
There have been various obstacles,  various significant achievements---and there are 
important projects yet to be done. I shall confine myself to reporting (from Holland 1993, 
Bohm and Hiley 1993) some salient points of work on:\\
\indent (i): single-particle relativistic wave equations: where I will emphasise the question 
whether sub-luminal trajectories can be defined from the current as usually defined; and\\
\indent (ii): quantum field theory: where I will give more details, and emphasise the use of 
a preferred frame and the non-locality of the quantum potential (Section \ref{qft}).

\indent So in various respects, what follows just scratches the surface of a large subject. 
For example, one topic which I will ignore is the pilot-wave account of many-particle wave 
equations; (for the many-particle Dirac equation, cf. Bohm and Hiley 1993, pp. 214-225, 
272-286; and briefly, Holland 1993, p. 509). So I will not discuss  guidance equations for 
particles  that are non-local on analogy with that in (iii) of Section \ref{nrpw}, viz. by 
taking a gradient of $S$ in configuration space at a point determined by all the particles' 
positions. But we will still see non-locality in Section \ref{qft} below, viz. in the 
behaviour of the quantum potential.

\noindent So first, I consider (i): single-particle wave equations. To give a pilot-wave 
interpretation of such equations, we naturally ask whether the formalism allows the 
definition of a 4-vector current $j^{\mu}$ with the properties  that:\\
\indent (a): $j^{\mu}$ is conserved, $\pl_{\mu}j^{\mu} = 0$; \\
\indent (b): $j^{\mu}$'s time-component $j^0$ is positive, and so might be interpreted as a 
probability density;\\
\indent (c): $j^{\mu}$ is always timelike, so that its integral curves can be the worldlines 
of the particle concerned, which therefore travels subluminally.

It turns out that the Klein-Gordon equation, describing a spin 0 particle, resists a 
pilot-wave interpretation. Its 4-vector current $j^{\mu}$ enjoys property (a), but violates 
(b) and (c). On the other hand, the Dirac equation has a current satisfying all of (a)-(c); 
(Holland 1993, pp. 498-509; Bohm and Hiley 1993, pp. 232-238 for  the Klein-Gordon equation, 
and pp. 214-222 for the Dirac equation).

\paragraph{Quantum field theory}\label{qft}
The pilot-wave approach in Section \ref{nrpw} carries over successfully to the relativistic 
quantum field theory of bosons, such as a spin 0 particle or the photon. The general idea is 
threefold:\\
\indent (a):  We first formulate the theory in the Schr\"{o}dinger picture, using the space 
representation of the field coordinate $\psi$. This formulation is entirely orthodox, though 
not widely presented by the textbooks.\\
\indent (b): Then we make the polar decomposition of the wave-function, obtaining (as in 
Section \ref{nrpw}) a conservation equation, and a quantum analogue of the Hamilton-Jacobi 
equation containing a new quantum potential term.\\
\indent (c):  Then we interpret (a) and (b), much as we did in Section \ref{nrpw}. Namely: 
the orthodox wave-function $\Psi$ is a physically real (though of course mathematically 
complex) field on the configuration space of the field $\psi$: i.e. it is a functional of the 
field $\psi$. $\Psi$ always evolves by the Schr\"{o}dinger equation. So far, again so 
orthodox; (at least as regards formalism---orthodoxy might cavil at calling a wave-function 
physically real). But there is also at all times an actual field configuration, which evolves 
by a guidance equation which is natural, derived from the formalism, and a generalization of 
that in Section \ref{nrpw}. And we recover the orthodox probabilistic results by averaging 
over field configurations, using the Born rule understood non-instrumentalistically.

\indent I shall say a bit more about (a)-(c), concentrating on the interpretative aspects in 
(c); and for simplicity, on a neutral, spin 0, massless particle, described classically by a 
{\em real} scalar function $\psi({\bf q}, t)$. (For many more details, cf. Holland 1993, pp. 
519-537; Bohm and Hiley 1993, pp. 238-247, 286-295.)

\noindent (a): {\em Space representation}:--- As usual in the transition from a quantum 
theory of a fixed number of particles to a quantum field theory:\\
\indent (i): the role of the coordinate $\bf q$ in the former, viz. being the value of a 
degree of freedom, is taken over  in the latter, by the field $\psi$; and:\\
\indent (ii): the role of the former's label $i$, viz. labelling the degrees of freedom, is 
taken over by the continuous index ${\bf x} \equiv {\bf q}$.\\
So the configuration space is the infinite-dimensional space of possible configurations 
$\psi: \mathR^3 \rightarrow \mathR$; and the wave-function is $\Psi[\psi({\bf x}), t] = 
\scp{\; \psi({\bf x})}{\Psi(t) \;}$. That is: the wave-function $\Psi$ is a functional of the 
real scalar field $\psi$, and a function of the time $t$: it is {\em not} a point-function of 
$\bf x$.\\
\indent $\Psi$ obeys the Schr\"{o}dinger equation $i \hbar \; \pl \Psi / \pl t = {\hat H} 
\Psi$; where the Hamiltonian operator  $\hat H$ is derived from the classical Hamiltonian by 
the usual canonical quantization heuristic that ``Poisson brackets become  commutators''; so 
that in a representation in which $\psi({\bf x})$ is diagonal, the momentum is given by the 
functional  derivative $-i \hbar \; \delta / \delta \psi$.

\noindent (b): {\em Quantum potential}:--- We make a polar decomposition of $\Psi$ as $\Psi = 
R \exp(iS / \hbar)$, with both $R = R[\psi({\bf x}), t]$ and $S = S[\psi({\bf x}), t]$ being 
real functionals of the field. Then the Schr\"{o}dinger equation yields: a conservation law 
suggesting that $R^2$ is a probability density for the field; and a quantum analogue of the 
classical field's Hamilton-Jacobi equation. This quantum analogue  adds to the classical 
equation a new potential term, $U[\psi, t] := - \frac{1}{2 R} \int d^3 x \;  \frac{\delta^2 
R}{\delta \psi^2}$. This is the field-theoretic quantum potential. Recalling the transition 
to quantum field theory summarized in (a) above, we see that it is analogous to the 
elementary quantum potential $U({\bf q}_i)$ for several particles labelled $i$: $ U({\bf 
q}_i) = - \frac{\hbar^2}{2m} \frac{1}{R} \; (\Sigma_i \; \nabla^2_i R )$; cf. (iii) of 
Section \ref{nrpw}. We also see that the field-theoretic quantum potential is inherently 
non-local.

\noindent (c): {\em Guidance equation, interpretation}:--- The field-theoretic guidance 
equation, determining the motion of the postulated actual field configuration, takes the 
general form $\frac{\pl \psi}{\pl t} = \frac{\delta S[\psi({\bf x}), t]}{\delta \psi({\bf 
x})}$. Recalling again the summary in (a) above, this is clearly analogous to Section 
\ref{nrpw}'s guidance equation ${\dot {\bf q}_i} = \frac{1}{m} \; \nabla_i S$. For a neutral, 
spin 0, massless particle, which would classically be governed by the wave equation $\Box 
\psi({\bf x}, t) = 0$, the guidance equation takes the form $\Box \psi({\bf x}, t) = - 
\frac{\delta U[\psi({\bf x}), t]}{\delta \psi({\bf x})}$.\\
\indent This guidance equation, and the Schr\"{o}dinger equation $i \hbar \; \pl \Psi / \pl t 
= {\hat H} \Psi$ from (a), are the fundamental equations of motion for the total system which 
comprises both an actual classical field configuration, and a wave-functional $\Psi$ on the 
space of such configurations. For our purposes, we need only two points about these two 
equations and their interpretation. The first concerns the approximate status of Lorentz 
symmetry; the second concerns the classical limit. (For more details about these points, cf. 
e.g. Holland (1993,  pp. 522-524).)

\indent (i):  Our use, from (a) onwards, of the parameter $t$  amounts to postulating an 
absolute simultaneity structure like that in Newtonian mechanics, Galilean quantum mechanics 
and the non-relativistic pilot-wave approach. In particular, the constructions summarized in 
(b) and (c) do {\em not} reveal $t$ to have been ``free up to'' the rotation of hyperplanes 
associated with a Lorentz boost: as happens for the time-evolution in orthodox relativistic 
quantum theories. Indeed, the guidance equation and the Schr\"{o}dinger equation are not 
Lorentz-covariant, and the actual field $\psi({\bf x}, t)$ is not a Lorentz scalar. But just 
as in the non-relativistic case, we recovered orthodox quantum mechanical results by 
averaging over possessed positions using $\mid \psi \mid^2$ as the probability density: so 
also here, one recovers the results of orthodox relativistic quantum theory by averaging. In 
particular, Lorentz-covariance is an emergent symmetry: it fails at the sub-quantum level of 
the field configurations, but holds good once we average over field configurations.

\indent (ii): For the pilot-wave approach, it is the right hand side term in the guidance 
equation, i.e. the ``quantum force'' (generalized gradient of a potential), which is 
responsible for the characteristic differences of the quantum theory from the classical 
theory. Accordingly, we expect to obtain the classical limit when the magnitude and gradient 
of the quantum potential are both negligible. And indeed, when this is so, the guidance 
equation reduces to the classical wave equation $\Box \psi({\bf x}, t) = 0$. But away from 
the classical limit, the evolution of the field $\psi({\bf x}, t)$ is in general highly 
non-local (and non-linear). That is: how the field changes here and now depends on the 
present value of the field arbitrarily far away.

We can now (at last!) summarize how this pilot-wave approach to relativistic  quantum 
theories violates relativistic causality. The situation is broadly like the 
action-at-a-distance in the non-relativistic theory, due to the guidance equation and the 
quantum potential. Namely, some fundamental equations of the theory are non-local with 
respect to the theory's absolute simultaneity structure; so that, setting aside doubts of 
Dickson's and Norton's sort (cf. Section \ref{aaad}), an individual process involves 
instantaneous causation as in Newtonian gravitation. Or in terms of the light-cone structure 
(which for the pilot-wave approach is emergent): individual processes involve spacelike 
causation. On the other hand, when we average over these individual processes, we recover the 
orthodox Lorentz-covariant formalism: in particular, micro-causality (Holland 1993, p. 523) 
and Section \ref{mink}'s two other formulations of relativistic causality---but now 
interpreted as ``mere'' ensemble statements.

Finally, Section \ref{qncontradns} raised the question how violations of relativistic 
causality avoid  contradictions, and announced that my examples would do so by forbidding a 
spacelike zig-zag, ``there and back'', into the causal past of an initial event---so that the 
usual ``bilking argument'' for a contradiction could not get started.

\indent For the pilot-wave approach, the situation is straightforward. In view of the 
underlying absolute simultaneity structure, there is obviously no threat of a zig-zag, or of 
a contradiction: no more than with the action-at-a-distance in Newtonian gravity. This point 
is made (along of course with other points about non-locality) by e.g. Holland, and Bohm and 
Hiley; (cf. Holland 1993, pp. 483-487, 494-495, 531-537; Bohm and Hiley 1993 pp. 286-287; 
Kaloyerou 1993, p. 337). And in Section \ref{three}'s discussion of the Drummond-Hathrell and 
Scharnhorst effects, the same point was made in a somewhat generalized form: namely, that for 
superluminal propagation to be free of  contradictions, it is obviously enough that there be 
one frame of reference in which all the propagation  is forward in time.

\subsection{The Newton-Wigner representation}\label{NW}
I turn to my second example, which I will treat much more briefly. In effect, it develops a 
point mentioned at the start of Section \ref{nrpw}:  that in orthodox non-relativistic 
quantum theory, wave-functions propagate instantaneously. The point now is: this sort of 
propagation  also occurs in a part of {\em conventional} relativistic quantum theory---namely 
in the Newton-Wigner representation. 

The Newton-Wigner representation applies equally well to a relativistic quantum theory of a 
fixed number of particles, and to a quantum field theory. It provides a basis of the (pure) 
state space (the wave-functions) consisting of strictly localized states. At first sight, 
this seems analogous to the Dirac delta-functions, or more generally wave-functions with 
compact spatial support, of the non-relativistic theory. But these states, and the spectral 
projectors of the corresponding position quantity, have some striking features.\\
\indent (i): The states propagate superluminally. Indeed: although there is no absolute 
simultaneity structure, they propagate instantaneously in the following sense. If at time $t 
= 0$ in some inertial frame, a Newton-Wigner wave-function $\psi(0)$ has compact spatial 
support (i.e. is non-zero only in a compact spacelike patch, $\Sigma$ say, of the $t= 0$ 
hyperplane), then at all later times $t > 0$, no matter how small, $\psi(t)$ is non-zero 
throughout all space. Agreed, the great majority of the amplitude lies in the future 
light-cone of $\Sigma$; and as $t$ grows, the percentage of this majority rapidly tends to 
100 percent. Nevertheless there is a ``superluminal tail''.\\ 
\indent (ii): Two spectral projectors associated with spacelike related spatial regions on 
two different spacelike hyperplanes will {\em not} commute.

These features seem to imply, respectively, that:\\
\indent (i'): One could signal superluminally by ``releasing'' a Newton-Wigner wave-packet 
initially confined to a compact region of space. (And here `signalling' could be taken in a 
strong sense, viz. as a non-vanishing probability of a detection, triggering some 
pre-arranged event, at spacelike separation.)\\
\indent (ii'): A (sharp, von-Neumann-style) non-selective measurement of one such projector 
can influence that statistics of the measurement of the other.

These features have been analysed, and the Newton-Wigner formalism much developed, especially 
by Fleming. Most recently, his (2003, 2004) include replies to recent literature; (the 
features are also surveyed in Fleming and Butterfield (1999, pp. 108-130, 153-162)). His work 
emphasises (among many other points) that:\\
\indent (a): Though these features, and the Newton-Wigner formalism, are little known,  they 
are {\em not} heterodoxies: they form part of the conventional framework  of relativistic 
quantum theories.\\
\indent (b): Various arguments can be given that (i') and (ii') are in fact {\em not} implied 
(and so causal loops are avoided). I will not go into these arguments: (Fleming and 
Butterfield (1999, p. 157) gives some references). Admittedly, they are partial, reflecting 
our lack of a developed theory of measurement for relativistic quantum theories. So here it 
must suffice to make three points.

First: the general theme, that violations of relativistic causality may well be indicative of 
future physics, also occurred in Section \ref{three}. Second: I stress that the ingredients 
of these partial arguments are very different from those in Sections \ref{qft} and 
\ref{three}. For example, one ingredient is to associate signalling with the group velocity 
of a wave (there being no superluminal 
group velocities in the Newton-Wigner representation); and another is to relativize the 
notion of localization to a hyperplane.

 Third: in view of Section \ref{qnconsensus}'s formulation of relativistic causality as 
micro-causality, i.e. commutativity of spacelike operators, I should stress two general 
points of ``reassurance'', about the non-commutation in (ii) above.\\
\indent (1): If the spatial regions associated with the projectors are on the {\em same} 
hyperplane, then the projectors are of course orthogonal, representing the fact that a single 
particle localized in one region has zero probability to be found in the other---and so 
commute.\\
\indent (2): The non-commutation in (ii) is {\em consistent} with micro-causality. For the 
expression of the Newton-Wigner operators, in terms of the operators that are the topic of 
micro-causality, involves an integral over an entire spacelike hyperplane. The 
non-commutation is thus ``explained'' by the existence of timelike paths connecting portions 
of the integrands in the integrals occurring in the definitions of both spectral projectors. 
(Of course, the fact that the Newton-Wigner representation and the conventional one are 
related by an unbounded integral means that the senses in which the Newton-Wigner operators, 
and the conventional ones, are ``associated'' with regions are distinct. Accordingly, 
elucidating these different senses is a main theme of the recent literature: cf. Fleming 
(2004, Sections 3c, 4d and 5b) and references therein.)

\section{References}\label{refs}
Adams, A., Arkani-Hamed, N., et al. (2006), `Causality, Analyticity and an IR Obstruction to 
UV Completion'. Available at: hep-th/0602178

Adler, S. (1971), `Photon Splitting and Photon Dispersion in a Strong Magnetic Field', {\em 
Annals of Physics (New York)} {\bf 67}, pp. 599-647.

Adler, S. (2003), `Why decoherence has not solved the measurement problem: a response to P.W. 
Anderson', {\em Studies in the History and Philosophy of Modern Physics} {\bf 34}, pp. 
135-142. 

Arntzenius, F. and Maudlin, T. (2005), `Time Travel and Modern Physics', in {\em The Stanford 
Encyclopedia of Philosophy} ed. E. Zalta. Available at:\\ 
http://www.seop.leeds.ac.uk/entries/time-travel-phys/

Barton, G.  (1990), `Faster than $c$ Light between Parallel Mirrors: the Scharnhorst effect 
rederived', {\em Physics Letters B} {\bf 237}, pp. 559-562. 

Barton, G. and Scharnhorst, K. (1993), `QED between Parallel Mirrors: light signals faster 
than $c$, or amplified by the vacuum', {\em Journal of Physics A: mathematical and general} 
{\bf 26}, pp. 2037-2046.

Beckman, D. et al. (2001), `Causal and localizable quantum operations', {\em Physical Review 
A} {\bf 64}, p. 052309. Available at: quant-ph/0102043  

Berkovitz. J. (2002), `On Causal Loops in the Quantum Realm', in Placek, T. and Butterfield, 
J., eds., {\em Non-locality and Modality},  Kluwer Academic (Nato Science Series, vol. 64), 
pp. 235-257.

Bohm, D. and Hiley, B. (1993), {\em The Undivided Universe}, London: Routledge. 

Born, M. (1962), {\em Einstein's Theory of Relativity}, New York: Dover.

Brown, H. (2005), {\em Physical Relativity}, Oxford: University Press.

Brown, H. and Pooley, O. (2001), `The Origin of the Spacetime Metric: Bell's `Lorentzian 
pedagogy' and its significance in general relativity',  in {\em Physics Meets Philosophy at 
the Planck Scale}, C. Callender and N. Huggett eds., Cambridge: University Press, pp. 
256-272.

Brown, H. and Pooley, O. (2006), `Minkowski space-time: a glorious non-entity', in {\em The 
Ontology of Spacetime}, ed. D. Dieks, Oxford: Elsevier. Available at: 
http://philsci-archive.pitt.edu/archive/00001661/

Bub, J. (1997), {\em Intrepreting the Quantum World}, Cambridge: University Press. 

Butterfield, J. (2001), `The End of Time?'. Available at: http://philsci-archive.pitt.edu/\\
archive/00000104/ and gr-qc/0103055. (A shortened version, without the passage cited in 
Section 4.1, appeared as: {\em British Journal for the Philosophy of Science} {\bf 53}, 2002, 
pp. 289-330.)

Butterfield, J. (2007), `Stochastic Einstein Locality Revisited', forthcoming in {\em British 
Journal for the Philosophy of Science}.

Dickson, M. (1998), {\em Quantum Chance and Non-Locality}, Cambridge: University Press.

Dittrich, W. and Gies, H. (1998), `Light propagation in non-trivial QED vacua', {\em Physical 
Review D} {\bf 58}, pp. 025004. Available at: hep-ph/9804375.

Dowe, P. (2000), {\em Physical Causation}, Cambridge: University Press.

Drummond, I. (2001), `Variable Light-Cone Theory of Gravity', {\em Physical Review} {\bf D 
63}, 043503. Available at: gr-qc/9908058. 

Drummond, I. and Hathrell, S.  (1980), `QED Vacuum Polarization in a  Background 
Gravitational Field, and its effect on the Velocity of Photons', {\em Physical Review D} {\bf 
22}, pp. 343-355.

Earman, J. (1986), {\em A Primer on Determinism}, Dordrecht: Reidel.

Earman, J. (1987), `Locality, Non-locality and Action-at-a-distance',   {\em Theoretical 
Physics in the 100 Years since Kelvin's Baltimore Lectures}, eds. P. Achinstein and R. 
Kargon, Cambridge, MA: MIT Press, pp. 449-490.

Earman, J. (1995), {\em Bangs, Crunches, Whimpers and Shrieks: singularities and acausalities 
in relativistic spacetimes}, Oxford: University Press.

Earman, J. (2004) `Determinism: What we have Learned and What we still don't Know', in  {\em 
Freedom and Determinism}, Topics in Contemporary Philosophy Series, vol. II, J.K. Campbell, 
M. O'Rourke and D. Shier (eds.), Seven Springs Press. Also available at: 
www.ucl.ac.uk/~uctytho/detearmanintro.html. 

Earman, J. (2006) `Determinism in Modern Physics', in {\em The Handbook of Philosophy of 
Physics}, eds. J. Earman and J. Butterfield, Amsterdam: Elsevier, pp. 1369-1434.

Fleming, G. (2003), `Observations on Hyperplanes: I State Reduction and Unitary Evolution', 
available at: http://philsci-archive.pitt.edu/archive/00001533/

Fleming, G. (2004), `Observations on Hyperplanes: II. Dynamical Variables and Localization 
Observables', available at: http://philsci-archive.pitt.edu/archive/00002085/

Fleming, G. and Butterfield, J. (1999), `Strange Positions', in {\em From Physics to 
Philosophy}, ed.s J. Butterfield and C. Pagonis, Cambridge: University Press, pp. 108-165.

Friedlander, F. (1975), {\em The Wave Equation on a Curved Spacetime}, Cambridge: University 
Press.

Geroch, R. and Horowitz, G. (1979), `Global structure of spacetimes', in {\em General 
Relativity: an Einstein Centennial Survey}, ed. S. Hawking and W. Israel, Cambridge 
University Press,  pp. 212-293.

Geroch, R. and Jang, P. (1975), `Motion of a Body in General Relativity', {\em Journal of 
Mathematical Physics} {\bf 16}, pp. 65-67.

Ghins, M. and Budden, T. (2001), `The Principle of Equivalence', {\em Studies in the History 
and Philosophy of Modern Physics} {\bf 32B}, pp. 33-52.

Hawking, S. and Ellis, G. (1973), {\em The Large-Scale Structure of Spacetime}  Cambridge: 
University Press.

Holland, P. (1993), {\em The Quantum Theory of Motion}, Cambridge: University Press.

Hollands, S. and Wald, R. (2001) `Local Wick polynomial and time ordered products of quantum 
fields in curved spacetime', {\em Communications in Mathematical Physics} {\bf 223}, pp. 
289-326. Available at: gr-qc/0103074.

Hollands, S. and Wald, R. (2002) `Existence of local covariant time ordered products of 
quantum fields in curved spacetime', {\em Communications in Mathematical Physics} {\bf 231}, 
pp. 309-345. Available at: gr-qc/0111108.

Hollowood, T. and Shore, G. (2007) `Violation of micro-causality in curved spacetime'. 
Available at: hep-th/0707.2302.

Hollowood, T. and Shore, G. (2007a) `The refractive index of curved spacetime; the fate of 
causality in QED'. Available at: hep-th/0707.2303.

Horuzhy, S. (1990) {\em Introduction to Algebraic Quantum Field Theory}, Kluwer Academic.

Kaloyerou, P. (1993) `The causal interpretation of the electromagnetic field: the EPR 
experiment', in {\em Bell's Theorem and the foundations of Modern Physics}, ed.s A van der 
Merwe, F. Selleri and G. Tarozzi, Singapore: World Scientific, pp. 315-337.

Kay, B. (1992), `The principle of locality and quantum field theory on (non globally 
hyperbolic) curved spacetimes', {\em Reviews in Mathematical Physics}, Special Issue, pp. 
167-195.

Lewis, D. (1973), `Causation', {\em Journal of Philosophy} {\bf 70}, pp. 556-567;  reprinted 
in his {\em Philosophical Papers}, volume II, Oxford: University Press, 1986, pp. 159-171.

Lewis, D. (1973a), {\em Counterfactuals}, Oxford: Basil Blackwell.

Lewis, D. (1976), `The Paradoxes of Time Travel', {\em American Philosophical Quarterly} {\bf 
13}, pp. 145-152; reprinted in his {\em Philosophical Papers}, volume II, Oxford: University 
Press, 1986, pp. 67-80; page reference to reprint.

Liberati, S., Sonego, S. and Viser M. (2002), `Faster-than-c Signals, Special Relativity and 
Causality', {\em Annals of Physics} {\bf 298}, pp. 167-185. Available at: gr-qc/0107091.

Mill, J.S. (1872), {\em A System of Logic: Ratiocinative and Inductive}, 8th edition, London: 
Longman, Green and Co., 1916.

Minkowski, H. (1908) `Space and time', in H. Lorentz, A. Einstein et al. {\em The Principle 
of Relativity}, New York: Dover (1923). 

Misner, C., Thorne, K. and Wheeler, J. (1973), {\em Gravitation}, San Francisco: W H Freeman.

Newton, I. (1692/93), `Four Letters to Richard Bentley', reprinted in ed. M. Munitz, {\em 
Theories of the Universe}, pp. 211-219, New York: Free Press, 1957.

Norton, J. (1985), `What was Einstein's Principle of Equivalence?', {\em Studies in the 
History and Philosophy of Science} {\bf 16}, pp. 203-246.
 
Norton, J. (2003),`Causation as Folk Science', {\em Philosophers' Imprint} {\bf 3} http:\\
//www.philosophersimprint.org/003004/; to be reprinted in H. Price and R. Corry, {\em 
Causation and the Constitution of Reality}, Oxford: University Press.

Norton, J. (2006), `Do the Causal Principles of Modern Physics Contradict Causal 
Anti-fundamentalism?', to appear in {\em Causality: Historical and Contemporary},
eds. P. K. Machamer and G. Wolters, University of Pittsburgh Press. Available at: 
http://philsci-archive.pitt.edu/archive/00002735/

Norton, J. (2006a), `The Dome: An Unexpectedly Simple Failure of Determinism', to appear in 
{\em Philosophy of Science}, Proceedings of 2006 PSA Meeting. Available at: 
http://philsci-archive.pitt.edu/archive/00002943/

Putnam, H. (1962), `It Ain't Necessarily So', {\em Journal of Philosophy} {\bf 59}, pp. 
658-670; reprinted in his {\em Mathematics, Matter and Method}, Cambridge: University Press, 
1975, pp. 237-249; page reference to reprint. 

Radzikowski, M. (1996), `Microlocal approach to the Hadamard condition in quantum field 
theory on curved spacetime', {\em Communications in Mathematical Physics} {\bf 179}, pp. 
529-553.

Redei, M. (2002), `Reichenbach's Common Cause Principle and Quantum Correlations', in {\em 
Modality, Probability and Bell's Theorems}, NATO Science Series II, vol. 64, eds. T. Placek 
and J. Butterfield, Dordrecht: Kluwer Academic; pp. 259-270.  

Redei, M., and Summers, S. (2002), `Local primitive Causality and the Common Cause Principle 
in Quantum Field Theory', {\em Foundations of Physics} {\bf 32}, pp. 335-355.

Rugh, S., Zinkernagel, H and Cao, T. (1999), `The Casimir Effect and the interpretation of 
the vacuum', {\em Studies in History and Philosophy of Modern Physics} {\bf 30B}, pp. 
111-139.

Salmon, W. (1984), {\em Scientific Explanation and the Causal Structure of the World}, 
Princeton: University Press.

Scharnhorst, K. (1990), `On Propagation of Light in the Vacuum between Plates', {\em Physics 
Letters B} {\bf 236}, pp. 354-359. 

Scharnhorst, K. (1998), `The Velocities of Light in Modified QED Vacua', {\em Annals of 
Physics (Leipzig)} {\bf 7}, pp. 700-709. Available at: hep-th/9810221.

Shore, G. (2003), `Quantum Gravitational Optics', {\em Contemporary Physics} {\bf 44}, pp. 
503-521. Available at: gr-qc/0304059.

Shore, G. (2003a), `Causality and Superluminal Light', Proceedings of a conference {\em Time 
and Matter}, at Venice 2002, pp. 45-66. Available at: gr-qc/0302116.

Shore, G. (2007), `Superluminality and UV Completion', Available at: hep-th/0701185.

Sorkin, R. (1993), `Impossible measurements on quantum fields', in {\em Directions in General 
Relativity}. eds. B.L. Hu and T.A. Jacobson, Cambridge: University Press. Available at: 
gr-qc/9302018

Steinmann, O. (2000) `Perturbative quantum electrodynamics and axiomatic
field theory', Texts and Monographs in Physics. Springer-Verlag,
Berlin.

Torretti, R. (1983), {\em Relativity and Geometry}, Oxford: Pergamon Press.

Wald, R. (1984), {\em General Relativity}, Chicago: University of Chicago Press.

Wald, R. (1994), {\em Quantum Field Theory in Curved spacetime and Black Hole 
Thermodynamics}, Chicago: University of Chicago Press.

Weinstein, S. (1996), `Strange Couplings and Spacetime Structure', {\em Philosophy of 
Science} {\bf 63} (Supplement: Proceedings), pp. S63-S70.

Weinstein, S. (2006), `Superluminal Signalling and Relativity', {\em Synthese} {\bf 148}, pp. 
381-399.

\end{document}